\documentclass[a4paper,11pt]{article}
\pdfoutput=1 
\usepackage{subfig}

\usepackage{jcappub} 

\usepackage[T1]{fontenc} 
\usepackage{siunitx}
\DeclareSIUnit \parsec {pc}
\DeclareSIUnit \year {yr}
\usepackage{amsmath}

\usepackage[normalem]{ulem}

\date{February 24, 2023}

\title{\boldmath LiteBIRD Science Goals and Forecasts. A Case Study of the Origin of Primordial Gravitational Waves using Large-Scale CMB Polarization.}

\author[1,2,3]{P.\,Campeti,}
\author[1,4]{E.\,Komatsu,}
\author[5,6,7]{C.\,Baccigalupi,}
\author[8,3,9]{M.\,Ballardini,}
\author[10,11,12]{N.\,Bartolo,}
\author[13,14]{A.\,Carones,}
\author[15]{J.\,Errard,}
\author[9,16]{F.\,Finelli,}
\author[17]{R.\,Flauger,}
\author[18]{S.\,Galli,}
% ...
\affiliation[1]{Max Planck Institute for Astrophysics, Karl-Schwarzschild-Str. 1, D-85748 Garching, Germany}
\affiliation[2]{Excellence Cluster ORIGINS, Boltzmannstr. 2, 85748 Garching, Germany}
\affiliation[3]{INFN Sezione di Ferrara, Via Saragat 1, 44122 Ferrara, Italy}
\affiliation[4]{Kavli Institute for the Physics and Mathematics of the Universe (Kavli IPMU, WPI), UTIAS, The University of Tokyo, Kashiwa, Chiba 277-8583, Japan}
\affiliation[5]{International School for Advanced Studies (SISSA), Via Bonomea 265, 34136, Trieste, Italy}
\affiliation[6]{INFN Sezione di Trieste, via Valerio 2, 34127 Trieste, Italy}
\affiliation[7]{IFPU, Via Beirut, 2, 34151 Grignano, Trieste, Italy}
\affiliation[8]{Dipartimento di Fisica e Scienze della Terra, Università di Ferrara, Via Saragat 1, 44122 Ferrara, Italy}
\affiliation[9]{INAF - OAS Bologna, via Piero Gobetti, 93/3, 40129 Bologna, Italy}
\affiliation[10]{Dipartimento di Fisica e Astronomia “G. Galilei”, Universita` degli Studi di Padova, via Marzolo 8, I-35131 Padova, Italy}
\affiliation[11]{INFN Sezione di Padova, via Marzolo 8, I-35131, Padova, Italy}
\affiliation[12]{INAF, Osservatorio Astronomico di Padova, Vicolo dell’Osservatorio 5, I-35122, Padova, Italy}
\affiliation[13]{Dipartimento di Fisica, Università di Roma Tor Vergata, Via della Ricerca Scientifica, 1, 00133, Roma, Italy}
\affiliation[14]{INFN Sezione di Roma2, Università di Roma Tor Vergata, via della Ricerca Scientifica, 1, 00133 Roma, Italy}
\affiliation[15]{Université de Paris, CNRS, Astroparticule et Cosmologie, F-75013 Paris, France}
\affiliation[16]{INFN Sezione di Bologna, Viale C. Berti Pichat, 6/2 – 40127 Bologna Italy}
\affiliation[17]{University of California, San Diego, Department of Physics, San Diego, CA 92093-0424, USA}
\affiliation[18]{Institut d'Astrophysique de Paris, CNRS/Sorbonne Université, Paris France}
% ...

\emailAdd{pcampeti@mpa-garching.mpg.de}
\emailAdd{komatsu@mpa-garching.mpg.de}

\abstract{We study the possibility of using the {\sl LiteBIRD} satellite $B$-mode survey to constrain models of inflation producing specific features in CMB angular power spectra. We explore a particular model example, i.e. spectator axion-SU(2) gauge field inflation. This model can source parity-violating gravitational waves
from the amplification of gauge field fluctuations driven by a pseudoscalar ``axionlike'' field, rolling for a few e-folds during inflation. The sourced gravitational waves can exceed the vacuum contribution at reionization bump scales by about an order of magnitude and can be comparable to the vacuum contribution at recombination bump scales. We argue that a satellite mission with full sky coverage and access to the reionization bump scales is necessary to understand the origin of the primordial gravitational wave signal and distinguish among two production mechanisms: quantum vacuum fluctuations of spacetime and matter sources during inflation. We present the expected constraints on model parameters from {\sl LiteBIRD} satellite simulations, which complement and expand previous studies in the literature. 
We find that {\sl LiteBIRD} will be able to exclude with high significance standard single-field slow-roll models, such as the Starobinsky model, if the true model is the axion-SU(2) model with a feature at CMB scales. 
We further investigate the possibility of using the parity-violating signature of the model, such as the $TB$ and $EB$ angular power spectra, to disentangle it from the standard single-field slow-roll scenario. We find that most of the discriminating power of {\sl LiteBIRD} will reside in $BB$ angular power spectra rather than in $TB$ and $EB$ correlations.}

\begin{document}
\maketitle
\flushbottom

\section{Introduction}
\label{sec:intro}

A stochastic background of primordial gravitational waves (hereafter GWs) predicted by the inflationary paradigm \cite{Grishchuk:1974ny, Starobinsky:1979ty} represents one of the main targets of ongoing experimental efforts in cosmology. 
The simplest models of inflation (realized by a single scalar field minimally coupled to gravity and slowly rolling down its potential) predicts several properties of the distributions of the scalar (density) fluctuations \cite{mukhanov/chibisov:1981, hawking:1982, starobinsky:1982, guth/pi:1982, bardeen/steinhardt/turner:1983}, which are in remarkable agreement with cosmological observations \cite{Komatsu:2014ioa, Planck:2018vyg, Planck:2018jri, Planck:2019kim}. 
Additionally, if the GW background is detected, it would provide definitive evidence for cosmic inflation \cite{Guth:1980zm, Sato:1980yn, Albrecht:1982wi, Linde:1981mu}.

The spectrum of inflationary gravitational waves (tensor modes) extends over about 21 decades in frequency and is measurable through several different means. 
Among them, the primordial $B$-modes of the cosmic microwave background (CMB) polarization represent the most promising probe to detect the inflationary stochastic GW background \cite{Kamionkowski:1996zd, Seljak:1996gy}, and also the closest in time. Other possible options include pulsar timing arrays (hereafter PTA) and laser interferometers (see, e.g. Ref.~\cite{Campeti:2020xwn} for a review). Even though recent PTA measurements presented strong evidence for the existence of a stochastic GW background \cite{NANOGrav:2023gor, NANOGrav:2023hvm, Antoniadis:2023rey, Antoniadis:2023zhi, Reardon:2023gzh, Xu:2023wog}, this signal is still compatible with an astrophysical origin, i.e., due to inspirals of supermassive black hole binaries \cite{Broadhurst:2023tus, Middleton:2020asl, NANOGrav:2020spf}. Nonetheless, the deviation of the observed signal from the expected astrophysical GW background could still have a primordial origin \cite{Unal:2023srk, Murai:2023gkv, Ellis:2023oxs}.

Whereas there has been no detection of the $B$ modes from primordial tensor modes yet, CMB experiments currently hold the tightest constraints on their amplitude, customarily parametrized through the tensor-to-scalar ratio $r$ parameter, i.e. the ratio of the amplitudes of the tensor and scalar primordial power spectra.  
An upper limit $r<0.036$ at 95\,\% C.L., established by the BICEP/Keck collaboration at the pivot scale of $k_0=0.05\,\mathrm{Mpc}^{-1}$ assuming a fixed cosmology \cite{BICEP:2021xfz}, was shown to increase to $r<0.042$ at 95\,\% C.L. when fitting also for $\Lambda$CDM parameters \cite{Tristram:2021tvh}. With the addition of \textit{Planck} PR3 data, the current upper limit reduces to $r<0.035$ at 95\,\% C.L. \cite{BICEP:2021xfz, Paoletti:2022anb}. An even tighter limit is obtained using \textit{Planck} PR4 data and likelihoods, together with BAO data, i.e. $r<0.032$ at 95\,\% C.L. \cite{Tristram:2021tvh}. 
Using a conditioned covariance matrix as advocated in Ref.~\cite{Beck:2022efr}, the upper limit from the same datasets was shown to increase to $r<0.037$ at 95\,\% C.L. for a profile likelihood approach \cite{Campeti:2022vom} and to $r<0.038$ at 95\,\% C.L. for a Monte Carlo Markov chain approach \cite{Beck:2022efr}. Finally, fitting also the slope of the tensor power spectrum (and adding the LIGO-Virgo KAGRA dataset to the previous ones), leads to an improved upper bound $r<0.028$ at 95\,\% C.L. at the pivot scale $k_0=0.01\,\mathrm{Mpc}^{-1}$ \cite{Galloni:2022mok}.
Given the importance of this measurement, several $B$-mode experiments, such as the ground-based Simons Array \cite{Westbrook:2018vod}, Simons Observatory (SO) \cite{SimonsObservatory:2018koc}, the South Pole Observatory \cite{Moncelsi:2020ppj}, the Cosmology Large Angular Scale Surveyor (CLASS) \cite{2016SPIE.9914E..1KH} and CMB-S4 \cite{CMB-S4:2016ple}, the balloon-borne SPIDER \cite{SPIDER:2021ncy} and the {\sl LiteBIRD} (Lite satellite for the study of $B$-mode polarization and Inflation from cosmic background Radiation Detection) satellite \cite{LiteBIRD:2022cnt}, are currently targeting this very faint primordial signal or will do so towards the end of this decade. 
 More specifically, {\sl LiteBIRD} is a strategic large-class mission selected by the Japan Aerospace Exploration Agency (JAXA) to be launched in the early 2030s. It will orbit the Sun-Earth Lagrangian point L2 and will map the CMB polarization over the entire sky for three years, using three telescopes in 15 frequency bands between 34 and 448 GHz.
{\sl LiteBIRD}, being a full-sky satellite mission, will also have access to the very largest scale $B$ modes produced during cosmic reionization, in addition to the smaller scale $B$ modes produced during cosmic recombination. Instead, planned ground-based CMB experiments will only be able to access the recombination signature at smaller scales.
With a detection limit at the level of $r\lesssim 10^{-3}$, {\sl LiteBIRD} and CMB-S4 are the most sensitive among the experiments that have already passed the proposal stage.

However, simply detecting $r$ on its own would not allow one to understand the origin of the primordial GW background. In the simplest scenario (i.e. standard single-field slow-roll inflation), the primordial scalar and tensor perturbations are produced by quantum vacuum fluctuations of the metric. The resulting GW background has three distinctive properties: $(i)$ a nearly scale-invariant spectrum (with a slight red tilt given by the inflationary consistency relation); $(ii)$ an almost Gaussian probability density function (pdf); and $(iii)$ no net circular polarization (i.e., non-chiral or parity conserving polarization). In this simple framework, $r$ can be directly related to the energy scale of inflation \cite{Lyth:1996im}. However, all previous properties characterizing the vacuum-produced GW background do not necessarily hold if matter sources (i.e. excited extra particle content) are active during inflation. Checking for the scale dependence of the tensor spectrum, the presence of non-Gaussianities in the tensor modes and the existence of parity-violating correlations, both at the CMB and interferometer scales, therefore becomes a necessary step before any claim on the origin of these tensor modes can be made \cite{Komatsu:2022nvu, Campeti:2022acx, Campeti:2020xwn, Campeti:2019ylm, Hiramatsu:2018nfa, Paoletti:2022anb}. Other models that can violate the above properties of vacuum-produced GW background are those introducing non-minimal coupling of the inflaton \cite{Lue:1998mq, Bartolo:2017szm, Bartolo:2018elp}, additional scalar fields \cite{Cook:2011hg, Senatore:2011sp, Biagetti:2013kwa, Carney:2012pk, Cai:2021yvq}, and modified gravity models \cite{LISACosmologyWorkingGroup:2022jok}. Non-zero spatial curvature and kinetically-dominated initial conditions for inflation could also play a role in amplification or suppression of the GW background \cite{Hergt:2022fxk, Hergt:2018crm}.

Model-building with additional matter sources typically poses a challenge: sources are always at least gravitationally coupled to the inflaton sector, which results also in excitation of non-Gaussian scalar modes \cite{Barnaby:2010vf, Mirbabayi:2014jqa, Ferreira:2014zia, Ozsoy:2014sba}, potentially clashing with the bounds from CMB data \cite{Namba:2015gja, Campeti:2022acx, Namba:2013kia}. In this context, two of the most successful categories of models are based on Abelian \cite{Sorbo:2011rz, Barnaby:2010vf, Barnaby:2012xt, Cook:2013xea, Cook:2011hg, Namba:2015gja, Shiraishi:2016yun, Domcke:2016bkh, Ozsoy:2020ccy, Choi:2015wva, Fujita:2018zbr, Kawasaki:2019hvt,Ozsoy:2020kat} and non-Abelian \cite{Maleknejad:2011jw, Maleknejad:2011sq, Maleknejad:2012fw, Maleknejad:2016qjz, Dimastrogiovanni:2012ew, Dimastrogiovanni:2016fuu, Obata:2016tmo, Agrawal:2017awz, Agrawal:2018mrg, Adshead:2013qp, Adshead:2013nka, Adshead:2016omu, Adshead:2017hnc,Watanabe:2020ctz} gauge fields. In this paper, we will focus on a model belonging to the latter category, which sources tensor modes from an SU(2) gauge field coupled to a pseudoscalar ``axionlike'' field through a Chern-Simons term \cite{Dimastrogiovanni:2016fuu}. Since the axion and gauge fields are spectators (i.e., their energy density is subdominant compared to that of the inflaton), inflation is achieved through a standard inflaton sector \cite{Dimastrogiovanni:2016fuu}. However, the Chern-Simons coupling breaks the conformal invariance of massless free gauge fields coupled to gravity, allowing for tachyonic amplification of gauge field perturbations, a process controlled by the speed of the axion rolling along its potential. Gauge field fluctuations, in turn, lead to a peak in the primordial tensor power spectrum, with a characteristic Gaussian bump shape \cite{Thorne:2017jft}.
This model is unique because it can source tensor modes from matter fields at linear order in the perturbed Einstein equations without breaking the statistical isotropy of the Universe. Therefore, unlike similar models such as axion-U(1) inflation \cite{Namba:2015gja, Campeti:2022acx}, which can source tensor modes only at second order in gauge-field perturbations, the axion-SU(2) model typically produces a negligible amount of sourced scalars and scalar non-Gaussianity compared to the vacuum-produced ones, allowing for sizeable amplitudes of a sourced GW background without spoiling agreement with current CMB bounds. On the other hand, the axion-SU(2) setup produces a strongly non-Gaussian tensor signal due to self-coupling of the gauge field, providing another distinctive (and potentially crucial) feature of this model compared to the inflationary paradigm \cite{Namba:2015gja, Shiraishi:2016yun, Agrawal:2017awz,Agrawal:2018mrg}. Generalizations to SU(N) gauge theories have also been considered and lead to the same phenomenology\cite{Fujita:2022fff}. 

From the perspective of future $B$-mode experiments, the axion-SU(2) setup appears to be a particularly interesting candidate to probe: it can source tensor modes exceeding the vacuum contribution by a factor of $\sim 5$ on the reionization bump scales in the CMB $B$-modes, while both contributions can still be comparable on the recombination bump scales \cite{Ishiwata:2021yne}.     
As we argue in section 2.5 of \cite{LiteBIRD:2022cnt}, this feature of the model highlights the benefits of a full-sky survey with access to the reionization bump, such as {\sl LiteBIRD}, when trying to distinguish between sourced and vacuum origins of the primordial GW background. Our motivation in this work is to investigate how well {\sl LiteBIRD} can test the properties $(i)$ and $(iii)$ of the vacuum GW background, i.e., the approximate scale-invariance of the spectrum and the parity symmetry. Specifically, we discuss whether {\sl LiteBIRD} observations can exclude standard single-field slow-roll inflation if tensors fluctuations arise from the production of matter in a way that breaks the approximate scale-invariance of the spectrum and produces non-zero parity-violating correlations. If observed data were found to be inconsistent with predictions of standard single-field slow-roll models, it would provide a strong motivation to test also the property $(ii)$, i.e., the Gaussianity of the stochastic GW background. The latter topic will be the subject of detailed investigation in a follow-up {\sl LiteBIRD} collaboration paper.

This work is part of a series of papers that present the science achievable by the {\sl LiteBIRD} space mission, expanding on the overview published in Ref.~\cite{LiteBIRD:2022cnt}. 
In particular, we expand on the discussion presented in section 2.5 of Ref.~\cite{LiteBIRD:2022cnt} with a more quantitative approach, using {\sl LiteBIRD} map-based simulations (including the component separation and angular power spectra estimation steps), to build a robust power spectrum covariance matrix and a likelihood based on the Hamimeche \& Lewis \cite{Hamimeche:2008ai} approximation. We then perform a parameter inference on a representative selection of parameter choices for the model, using a frequentist Monte Carlo approach based on the Feldman-Cousins prescription \cite{Feldman:1997qc, ParticleDataGroup:2020ssz} (see also \cite{SPIDER:2021ncy, Herold:2021ksg, Campeti:2022vom, Campeti:2022acx, planck_profile} for recent applications in cosmology) to account for the presence of physical boundaries on the parameters. 

We also investigate another unique signature of the axion-SU(2) model: $TB$ and $EB$ parity-violating correlations in the CMB due to the Chern-Simons coupling. We find that the $TB$ and $EB$ spectra produced by this model cannot be detected by {\sl LiteBIRD}, and that almost all the constraining power on the axion-SU(2) model resides in the $BB$ spectrum.
We finally assess the power of {\sl LiteBIRD} in discriminating between the two possible mechanisms for the production of gravitational waves, including information from $TT$, $EE$, $BB$, $TE$, $TB$ and $EB$ CMB spectra in the covariance matrix. 

The rest of the paper is structured as follows. In section \ref{sec:theory} we review the axion-SU(2) model of \cite{Dimastrogiovanni:2016fuu} and the latest bounds available in the literature. In section \ref{sec:method} we describe the simulations and the method used. In section \ref{sec:wrong_fit} we emphasize the benefits of a full-sky $B$-mode mission to test the origin of the stochastic GW background. In section \ref{sec:constraints} we quantify the constraining power of {\sl LiteBIRD} on the parameters of the axion-SU(2) model. In section \ref{sec:TBEB} we investigate the additional discriminating power of the $TB$ and $EB$ angular power spectra. We present our conclusions in section \ref{sec:conclusions}.

\section{Spectator axion-SU(2) gauge field inflation model}\label{sec:theory}

We will consider the spectator axion-SU(2) gauge field inflation model of Ref.~\cite{Dimastrogiovanni:2016fuu}, based on the ``chromo-natural'' inflation model \cite{adshead/wyman:2012} (see also Refs.~\cite{Komatsu:2022nvu, Maleknejad:2012fw} for reviews).  The Lagrangian for this model,   
\begin{equation}
\label{eq:action}
    \mathcal{L}=\mathcal{L}_{\text{inf}}-\frac{1}{2}\left(\partial_{\mu}\chi\right)^{2}-V(\chi)-\frac{1}{4}F_{\mu\nu}^{a}F^{a\mu\nu}+\frac{\lambda}{4f}\chi F_{\mu\nu}^{a}\Tilde{F}^{a\mu\nu},
\end{equation}
contains a standard inflaton sector $\mathcal{L}_{\text{inf}}$, which always dominates the energy density and is responsible for inflation, and an axion field $\chi$ with a cosine-type potential $V(\chi) = \mu^{4}\left[1+\cos\left( \chi / f\right)\right]$, where $\mu$ and $f$ are dimensioned parameters and $\lambda$ is the dimensionless coupling constant for an SU(2) gauge field coupled to the axion via a Chern-Simons term.  The SU(2) gauge field $\mathbf{A}_{\mu}=\sum_a A_{\mu}^a\mathbf{\sigma}_a$ (where $\mathbf{\sigma}_a$ are the Pauli matrices and $a=\{1,2,3\}$) has a field strength tensor given by $F_{\mu\nu}^{a}=\partial_{\mu}A^{a}_{\nu}-\partial_{\nu}A^{a}_{\mu}-g \epsilon^{abc}A^{b}_{\mu}A^{c}_{\nu}$, where $g$ is the self-coupling constant, $\Tilde{F}^{a\mu\nu}\equiv \epsilon^{\mu\nu\rho\sigma}F_{\rho\sigma}^{a}/(2\sqrt{-g_{\rm M}})$ is the dual field strength tensor and $\epsilon^{\mu\nu\rho\sigma}$ is the totally antisymmetric symbol with $\epsilon^{0123}=1$. Here, $\sqrt{-g_{\rm M}}$ is the determinant of the metric tensor.

This model can source tensor modes at linear order in the perturbed Einstein equations without breaking the observed statistical isotropy of the Universe, because the SU(2) gauge field establishes a homogeneous and isotropic background solution, 
$\bar A_i^a=a(t)Q(t)\delta^a_i$ \cite{Maleknejad:2011sq,Maleknejad:2011jw}. This solution is approached even if the Universe was initially highly anisotropic (i.e. is an attractor solution) \cite{maleknejad/erfani:2014,domcke/etal:2019,wolfson/maleknejad/komatsu:2020, Wolfson:2021fya}. The perturbation around this solution gives scalar, vector, and tensor modes \cite{Maleknejad:2011sq,Maleknejad:2011jw}. In particular, tensor fluctuations of the gauge field, which are subject to tachyonic amplification near horizon crossing, source gravitons at the linear level in the stress-energy tensor. Only one of the helicities of the gauge field is amplified due to the parity-violating Chern-Simons coupling, leading to a \textit{chiral} GW background with left- or right-handed circular polarization \cite{Adshead:2013qp, Adshead:2013nka, Dimastrogiovanni:2012ew,Maleknejad:2012fw}. The primordial power spectrum of the sourced tensor modes, assuming that only left-handed gravitational waves are amplified, has a log-normal shape\footnote{Note that this result assumes also that the slow-roll approximation is appropriate for the axion evolution. This has been checked in Ref.~\cite{Thorne:2017jft} comparing with the full numerical solution of the background and perturbation equations for the axion and gauge field.} \cite{Thorne:2017jft}
   \begin{align}\label{eq:template}
        \mathcal{P}_{\rm t}^{L,\, {\rm sourced}} (k) &= r_{*} \mathcal{P}_{\zeta}(k_{\rm p}) \exp\left[-\frac{1}{2\sigma^{2}} \ln^{2}\left(\frac{k}{k_{\rm p}}\right) \right], \\
        \mathcal{P}_{\rm t}^{R,\, {\rm sourced}} (k) &\simeq 0,
    \end{align}
 controlled by the wavenumber $k_{\rm p}$ where the spectrum peaks, the effective tensor-to-scalar ratio at the peak scale $r_*$, the width of the Gaussian-shaped bump $\sigma$ and the power spectrum of scalar curvature perturbations $\mathcal{P}_{\zeta}$. The three parameters $\left\{r_{*}, k_{\rm p}, \sigma \right\}$ can be related to the parameters in the model Lagrangian\footnote{The three parameters $\left\{r_{*}, k_{\rm p}, \sigma \right\}$ do not uniquely determine the four Lagrangian parameters $\left\{g, \lambda, \mu, f \right\}$: it is necessary to specify a fourth parameter, e.g. $m_*$, the mass of the gauge field fluctuations at the inflection point of the potential  \cite{Thorne:2017jft}.} $\left\{g, \lambda, \mu, f \right\}$ (Eq.~\ref{eq:action}) \cite{Thorne:2017jft,fujita/sfakianakis/shiraishi:2019}.
The peak wavenumber $k_{\rm p}$ corresponds to the time $t_*$ at which $\chi$ is at the inflection point of the potential, $\chi(t_*)=\pi f/2$, reaching its maximum velocity. We also define the dimensionless time-dependent mass parameter of the gauge field fluctuations as $m_{Q}(t)\equiv gQ(t)/H$ (with $H$ the Hubble expansion rate during inflation), which is $m_*\equiv m_Q(t_*)=(g^2\mu^4/3\lambda H^4)^{1/3}$ at the inflection point. We can also define a dimensionless effective coupling $\xi_*\equiv \lambda\dot\chi(t_*)/(2fH)\approx m_*+m_*^{-1}$ and write
$k/k_{\rm p}=e^{H(t-t_*)}$, $\sigma^2=(\lambda/2\xi_*)^2/[2{\cal G}(m_*)]$, with ${\cal G}(m_*)\approx 0.666+0.81m_*-0.0145m_*^2-0.0064m_*^3$ \cite{Thorne:2017jft}. The effective tensor-to-scalar ratio at the peak scale can be defined as 
\begin{equation}
r_*=\frac{\mathcal{P}_{\rm t}^{\rm sourced}(k_{\rm p})}{\mathcal{P}_{\zeta}(k_{\rm p})} = \frac{m_{*}^4 H^{4}\mathcal{F}(m_*)}{\pi^2 g^2 M_{\rm Pl}^4 \mathcal{P}_{\zeta}(k_{\rm p})},    
\end{equation}
where $\mathcal{F}(m_Q)\simeq \exp{[2.4308m_Q - 0.0218m_Q^2 - 0.0064m_Q^3-0.86]}$ and $M_{\rm Pl}$ is the Planck mass, and can take any positive value, in principle\footnote{Note that the expression for $\mathcal{F}(m_Q)$ is valid for $3 \leq m_Q \leq 7$ \cite{Thorne:2017jft}.}. On the other hand, since $\chi$ can remain at the top of its cosine-type potential hill only for a limited amount of time due to its quantum fluctuations, $\sigma$ and $k_{\rm p}$ must satisfy the relation \cite{Thorne:2017jft}: 
\begin{equation}
   \Delta N = \sigma \sqrt{2\mathcal{G}(m_*)} \gtrsim \Delta N_{\mathrm{min}} = \frac{1}{1.8} \log \left(\frac{k_\mathrm{p}}{k_{\mathrm{CMB}}}\right),  
\end{equation}
where $\Delta N$ is the number of $e$-folds during which the axion is rolling down its potential
and $k_{\text{CMB}}$ is roughly equal to the largest observable CMB scale. 
Whereas Eq.~\ref{eq:template} and the following discussion hold only for axion potentials of the cosine-type and those with an inflection point, other power spectrum shapes are possible for different $V(\chi)$ \cite{fujita/sfakianakis/shiraishi:2019}.

In all previous equations, $\mathcal{P}_{\zeta}$ receives negligible sourced contributions for $m_{Q}\geq \sqrt{2}$ \citep{Dimastrogiovanni:2012ew,Dimastrogiovanni:2016fuu} and can therefore be assumed to be equal to the vacuum scalar power spectrum ${\mathcal P}^{\rm vac}_{{\mathcal R}}(k)= A_{\rm s} \left(k / k_0\right)^{n_{{\rm s}}-1}\,$, with the amplitude $A_{\rm s}$, spectral index $n_{\rm s}$ and pivot scale $k_{0}=0.05\,\SI{}{\per\mega\parsec}$. 
The tensor power spectrum from the inflaton sector (indicated by ``vac'') receives instead a sourced contribution:
      \begin{align}
       \mathcal{P}_{\rm t}(k, k_{\rm p}, r_{*}, \sigma) &= \mathcal{P}_{\rm t}^{\rm vac}(k) + \mathcal{P}_{\rm t}^{\rm sourced}(k, k_{\rm p}, r_{*}, \sigma),\\
    \mathcal{P}_{\rm t}^{\rm sourced}(k, k_{\rm p}, r_{*}, \sigma) &= \mathcal{P}_{\rm t}^{L,\, {\rm sourced}} (k) + \mathcal{P}_{\rm t}^{R,\, {\rm sourced}} (k),
   \end{align}
where  ${\mathcal P}^{\rm vac}_{\rm t}(k)= A_{\rm t} \left(k/k_0\right)^{n_{\rm t}}$ is the tensor power spectrum from quantum vacuum fluctuations, with the amplitude $A_{\rm t}$ and spectral index $n_{\rm t}$. We also define the tensor-to-scalar ratio of purely vacuum fluctuations as $r_{\rm vac}=A_{\rm t}/A_{\rm s}$.    

\begin{figure}
    \centering
    \includegraphics[scale=0.5]{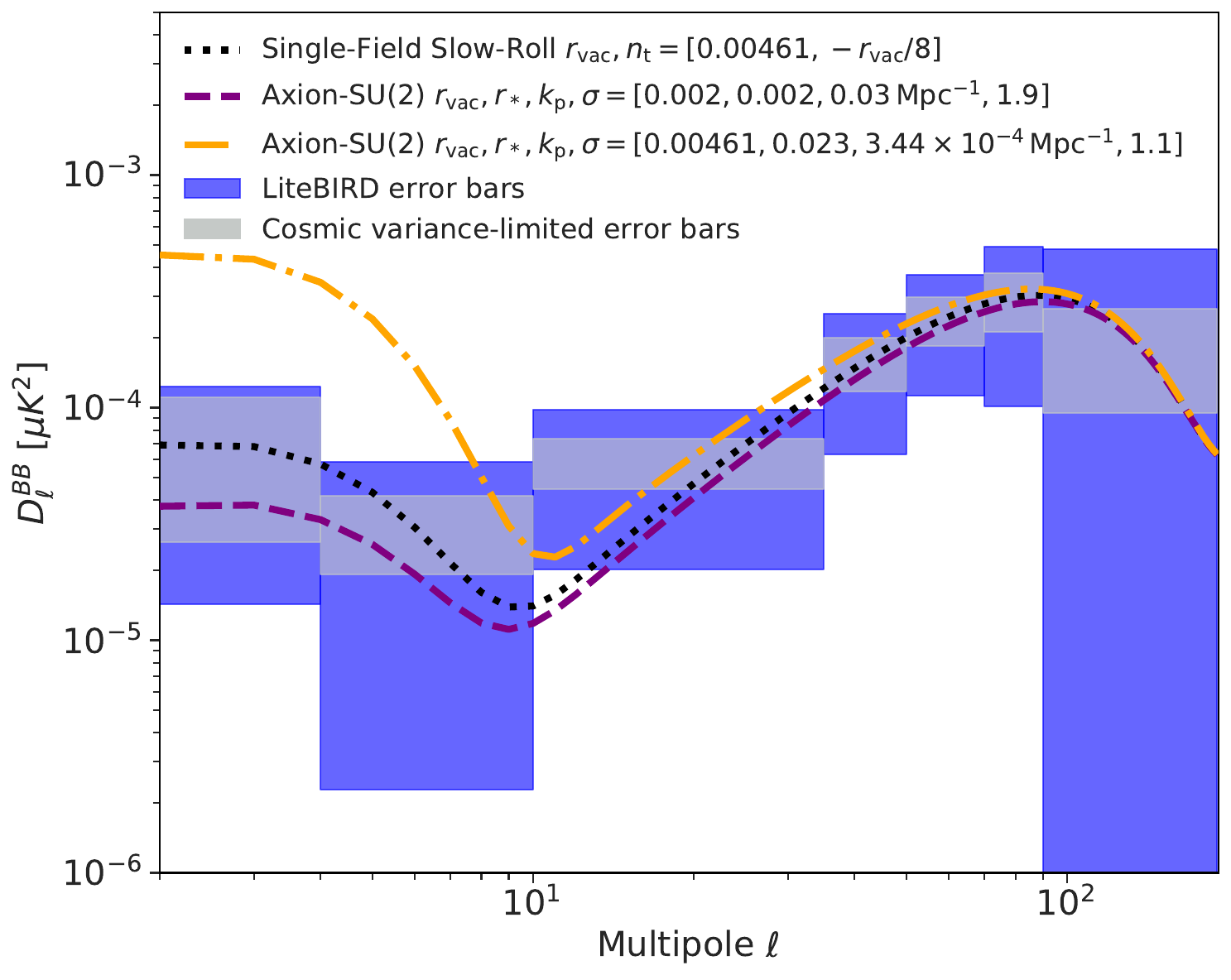}
    \caption{$B$-mode power spectra, $D_{\ell}^{BB} = \ell(\ell + 1)C_{\ell}^{BB}/2\pi$, for the Starobinsky model with $r_{\rm vac}=0.00461$ and $n_{\rm t}=-r_{\rm vac}/8$ (black dotted line) and for axion-SU(2) inflation with two parameter sets (see section \ref{sec:theory}): one gives the ``high reionization bump'' model (dash-dotted orange) with parameters $r_*, \sigma, k_{\rm p}, r_{\rm vac}=[0.023, 1.1, 3.44\times10^{-4}\,\mathrm{Mpc}^{-1}, 0.00461]$; the other the ``low reionization bump'' model (dashed purple) with $r_*, \sigma, k_{\rm p}, r_{\rm vac}=[0.002, 1.9, 0.03\,\mathrm{Mpc}^{-1}, 0.002]$. The cosmic-variance-only (including primordial and lensing $B$-mode variance) and total {\sl LiteBIRD} $\pm 1\,\sigma$ binned error bars (including foreground residuals) are shown as the gray and blue regions, respectively.}
    \label{fig:ptep_updated}
\end{figure}

The theoretical self-consistency of the axion-SU(2) setup has been studied in a number of papers, focusing mainly on the backreaction effect of particle production from the background axion and gauge fields on the background evolution, which could possibly ruin the phenomenological success of this model \cite{Maleknejad:2018nxz,Lozanov:2018kpk, Domcke:2018gfr,Mirzagholi:2019jeb, Maleknejad:2019hdr, Fujita:2017jwq, Dimastrogiovanni:2016fuu, Ishiwata:2021yne,Iarygina:2023mtj}. Recently, the study on spin\nobreakdash-2 particle production has been significantly improved by solving equations of motion with backreaction for a wide range of model parameters \cite{Ishiwata:2021yne}. According to this study, the amplitude of the sourced tensor modes can exceed by more than a factor $\mathcal{O}(10)$ the vacuum contribution at the CMB scales, and by several orders of magnitude at smaller scales. Including the effect of non-Gaussian scalar perturbations produced in second order by sourced tensor modes \cite{Papageorgiou:2018rfx, Papageorgiou:2019ecb} reduces the allowed ratio of sourced-to-vacuum tensors to a factor $\mathcal{O}(1)$ at $\ell\gtrsim 80$, where both the scalar power spectrum and scalar non-Gaussianity are tightly constrained by CMB temperature data \cite{Planck:2018vyg, Planck:2019kim, Komatsu:2010hc}, and to a factor $\sim5$ at low multipoles $\ell\lesssim 10$, where the temperature constraints are weaker.   
 
In light of these theoretical bounds, we choose two sets of parameters of the axion-SU(2) model with the purpose of obtaining $B$-mode spectra with reionization bump scales ($\ell\lesssim 10$) \textit{as different as possible from the Starobinsky model} of inflation \cite{starobinsky:1980} ($r_{\rm vac}=0.00461$, $n_{\rm t}=-r_{\rm vac}/8$), \textit{while having similar recombination bumps} ($\ell \sim 80-100$) (Fig.~\ref{fig:ptep_updated}). These two sets of parameters are $r_*, \sigma, k_{\rm p}, r_{\rm vac}=[0.023, 1.1, 3.44\times10^{-4}\,\mathrm{Mpc}^{-1}, 0.00461]$, which gives the ``high reionization bump'' model (dot-dashed orange curve in Fig.~\ref{fig:ptep_updated}) and $r_*, \sigma, k_{\rm p}, r_{\rm vac}=[0.002, 1.9, 0.03\,\mathrm{Mpc}^{-1}, 0.002]$, the ``low reionization bump'' model (dashed purple). These two choices update those in figure 4 of Ref.~\cite{LiteBIRD:2022cnt} with viable models according to the study in Ref.~\cite{Ishiwata:2021yne}. The CMB angular power spectrum for the ``high reionization bump'' remains approximately the same as in Ref.~\cite{LiteBIRD:2022cnt}, although we use $r_{\rm vac}=0.00461$ instead of the almost negligible value of $r_{\rm vac}=10^{-4}$ previously assumed. This is because the ratio of sourced-to-vacuum tensors can be at most $\sim5$ at low multipoles \cite{Ishiwata:2021yne}. In this new choice of parameters, the vacuum fluctuations provide a similar recombination bump as the Starobinsky model, while the sourced modes enhance only the reionization bump. However, now a value of $r_* = 0.023$ (about half of the previous $r_* = 0.041$) is sufficient to obtain the amplitude of the power spectrum similar to the one in Ref.~\cite{LiteBIRD:2022cnt} at scales $\ell\lesssim 10$. On the other hand, the new ``low reionization bump'' model, due to the fact that the ratio of sourced-to-vacuum tensors can be at most $\mathcal{O}(1)$ at $\ell\gtrsim 80$ \cite{Ishiwata:2021yne}, is significantly different from the one in \cite{LiteBIRD:2022cnt}. In this case, both vacuum and sourced tensors contribute equally (i.e., $r_{\rm vac}= r_{*}$) to produce a recombination bump similar to that of the Starobinsky model. Figure \ref{fig:ptep_updated} also shows for reference the cosmic variance-only (including primordial and lensing $B$-mode variance) and total {\sl LiteBIRD} $\pm 1\,\sigma$ binned error bars (including foreground residuals) as gray and blue regions, respectively. 

In this work, we also check that all the parameter choices are consistent with observational constraints on the axion-SU(2) model from the analysis of current CMB datasets \cite{Hamann:2022apw}. However, as also noted in \cite{Hamann:2022apw}, the shape of the tensor power spectrum is very weakly constrained by the \textit{Planck} and BICEP/Keck data: the upper limits on the model parameters are prior dominated and strongly affected by degeneracies.

\section{Method, simulations and likelihood}\label{sec:method}

\begin{figure}
    \centering
    \includegraphics[scale=0.225]{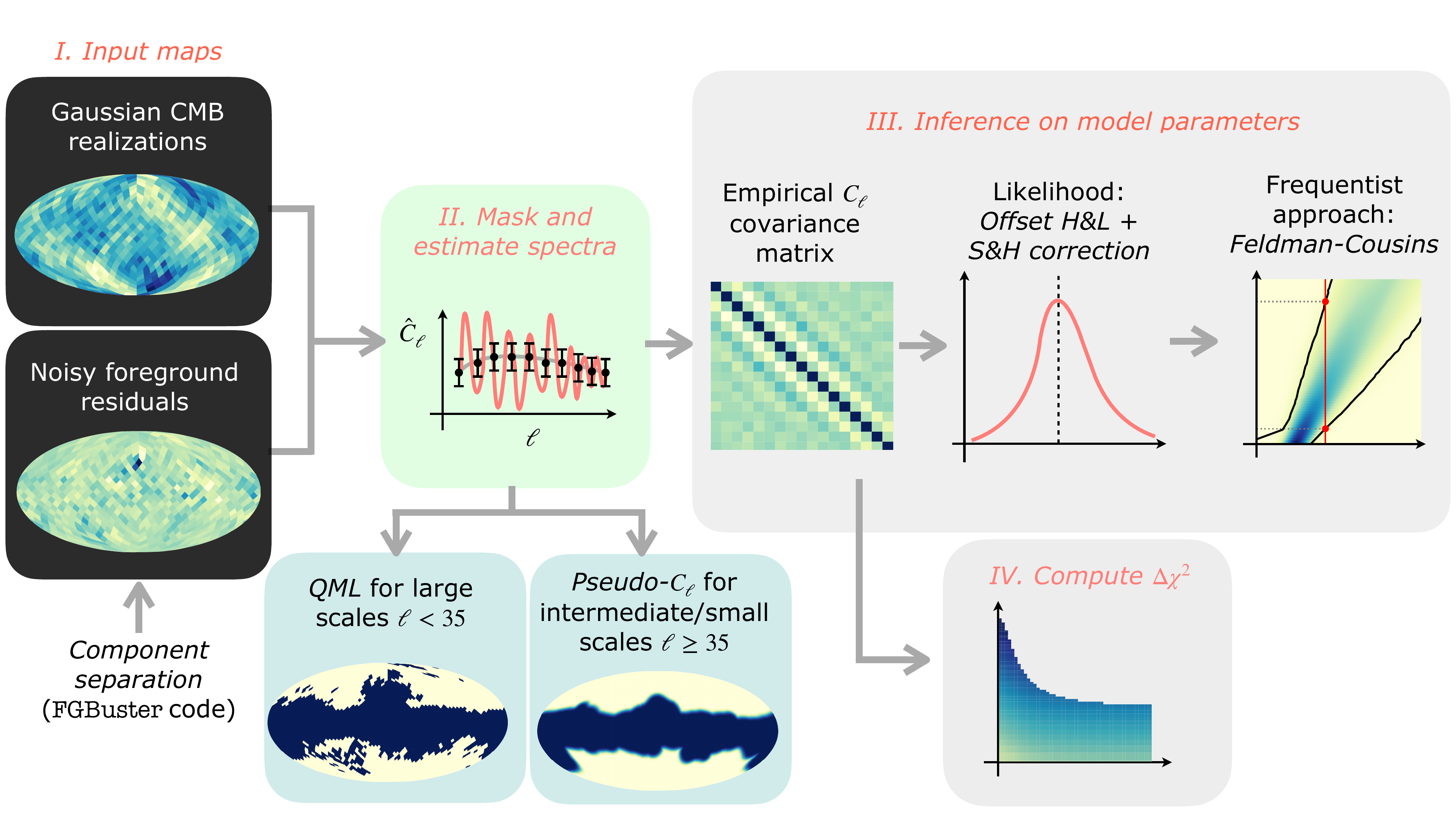}
    \caption{Pipeline for inflationary model parameter constraints used in the paper. See sections \ref{sec:method}, \ref{sec:constraints} and \ref{sec:TBEB} for details.}
    \label{fig:map}
\end{figure}

In this section, we will briefly describe the map-based simulations with {\sl LiteBIRD} specifications, the component separation method, the angular power spectrum estimation procedure, and the likelihood used in this work. 
A sketch of the pipeline from input maps to inference on inflationary model parameters is shown in Fig.~\ref{fig:map}. 

\medskip
\noindent{\bf Simulations.} The simulation setup used in this paper is the same as described in section 5.2 of \cite{LiteBIRD:2022cnt} to which we refer the reader for full details. Here, we summarize the main points. The input CMB and Galactic foreground maps in each {\sl LiteBIRD} frequency channel are generated at a \texttt{HEALPix} \cite{Gorski:2004by} resolution of $N_{\text{side}}=512$. The input maps of the CMB are generated as random Gaussian realizations from the lensed CMB angular power spectra corresponding to the values of the \textit{ Planck} best-fit cosmological parameters \cite{Planck:2018vyg} and $r_{\rm vac}=0$. Note that simulations for $r_{\rm vac} \neq 0$ and/or other models of inflation are built from the component separation outputs of the maps with $r_{\rm vac}=0$ as explained in the ``Component separation'' section below. The input maps of the Galactic foreground are instead generated using the \texttt{PySM} package \cite{Thorne:2016ifb, Zonca:2021row}, and more specifically the default model identified as \texttt{d1s1}, which includes templates of thermal dust and synchrotron emission \cite{Planck:2015mvg, Remazeilles:2014mba, WMAP:2012fli}. The frequency dependence of the thermal dust signal is modeled as a modified black-body with temperature and spectral index varying across the sky, following \cite{Planck:2015mvg}. Similarly, polarized synchrotron emission follows a power law with a spatially varying spectral index \cite{Miville-Deschenes:2008lza}. 
The sky templates from \texttt{PySM} are then convolved with top-hat bandpasses, coadded, convolved with Gaussian circular beams, and subsequently added to instrumental (white) noise realizations (see Table 3 in Ref.~\cite{LiteBIRD:2022cnt} for bandpass width, beam size, and noise levels considered). This procedure results in 1000 realizations of CMB, noise and foregrounds, which constitute the input maps for the component separation procedure described below.

\medskip
\noindent{\bf Component separation.} Component separation is performed using the \texttt{FGBuster}\footnote{\url{https://github.com/fgbuster/fgbuster}} code \cite{Stompor:2008sf}, implementing  a parametric fitting approach. The foreground model assumed in the code includes three spectral parameters: the spectral index and temperature of a modified black-body spectrum for the dust; and the spectral index of the power-law spectrum of synchrotron radiation. The three spectral parameters are fitted independently over patches defined by \texttt{HEALPix} pixels at a given $N_{\text{side}}$ resolution \cite{Errard:2018ctl}. To find a suitable balance between the statistical uncertainty on spectral parameters, which increases with the number of pixels (i.e., of free parameters in the fit), and the need to capture the actual spatial variability of the foreground sky, we adopted different values of $N_{\text{side}}$ for each parameter and according to the Galactic latitude \cite{LiteBIRD:2022cnt}. 

The \texttt{FGBuster} code returns a map including post-component separation noise and foreground residuals (hereafter collectively named ``noisy foreground residuals'') at $N_{\text{side}}=64$ for each of the 1000 input CMB and noise map realizations. Since these output maps are independent of the CMB input signal \cite{Stompor:2016hhw}, we can sum them to Gaussian map realizations of a lensed CMB signal computed from an arbitrary cosmological model. In this way, we can get 1000 component-separated maps for each choice of our inflationary model parameters. This feature will be used in sections \ref{sec:wrong_fit}, \ref{sec:constraints} and \ref{sec:TBEB} to obtain constraints on the axion-SU(2) model without rerunning the computationally expensive component-separation code for each set of model parameters. To avoid correlations, we used 500 simulations to estimate the $C_{\ell}$ covariance matrix with a fiducial model and the other 500 simulations as simulated data (see also below).

\medskip
\noindent{\bf Angular power spectrum estimation.}
We compute auto-spectra of maps using two different estimators of angular power spectra at different scales, as in Ref.~\cite{Tristram:2020wbi}: a pure pseudo-$C_{\ell}$s method (implemented in the \texttt{NaMaster} \footnote{\url{https://github.com/LSSTDESC/NaMaster}} code \cite{Alonso:2018jzx}) for intermediate and small scales ($\ell\geq 35$) and a quadratic maximum likelihood (QML) estimator \cite{Tegmark:2001zv, Vanneste:2018azc} at large scales ($\ell<35$). The pseudo-$C_{\ell}$s estimator is nearly optimal for smaller scales, but suboptimal at large scales \cite{Grain:2009wq, Grain:2012cx}. The QML method, despite being more computationally expensive compared to the pseudo-$C_{\ell}$s, produces nearly optimal variance estimates of the power spectrum at large scales. We use the QML implementation publicly available in the \texttt{xQML}\footnote{\url{https://gitlab.in2p3.fr/xQML/xQML}} code \cite{Vanneste:2018azc}. The two estimators are computed separately on the maps and then joined to form a single auto power spectrum.
As specified above, the noisy foreground residuals returned by \texttt{FGBuster} have a resolution of $N_{\text{side}}=64$; however, since the \texttt{xQML} implementation is memory intensive and therefore not feasible at this resolution, we downgrade the maps to $N_{\text{side}}=16$, after convolution with an apodizing kernel to avoid aliasing \cite{Planck:2018vyg}.

\begin{figure}
  \centering
  \subfloat{\includegraphics[width=0.49\textwidth]{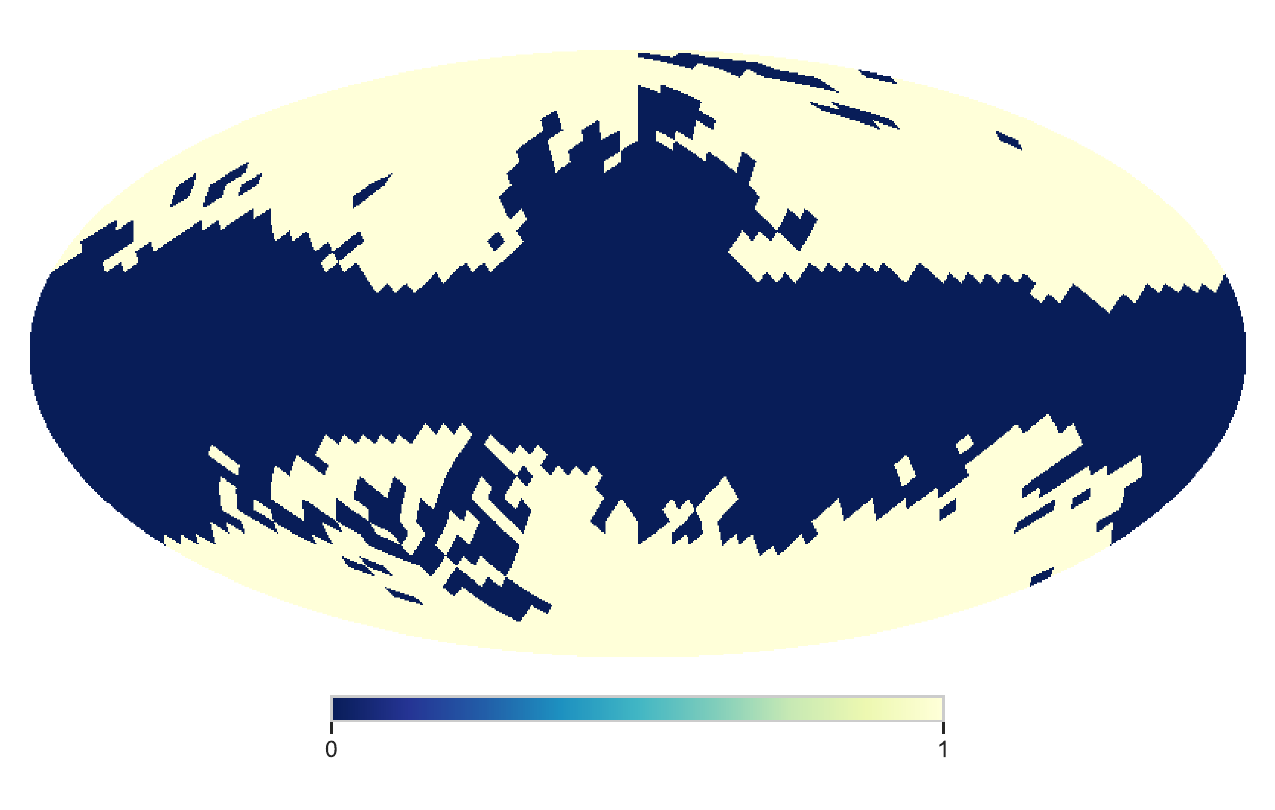}\label{fig:f1}}
  \hfill
  \subfloat{\includegraphics[width=0.49\textwidth]{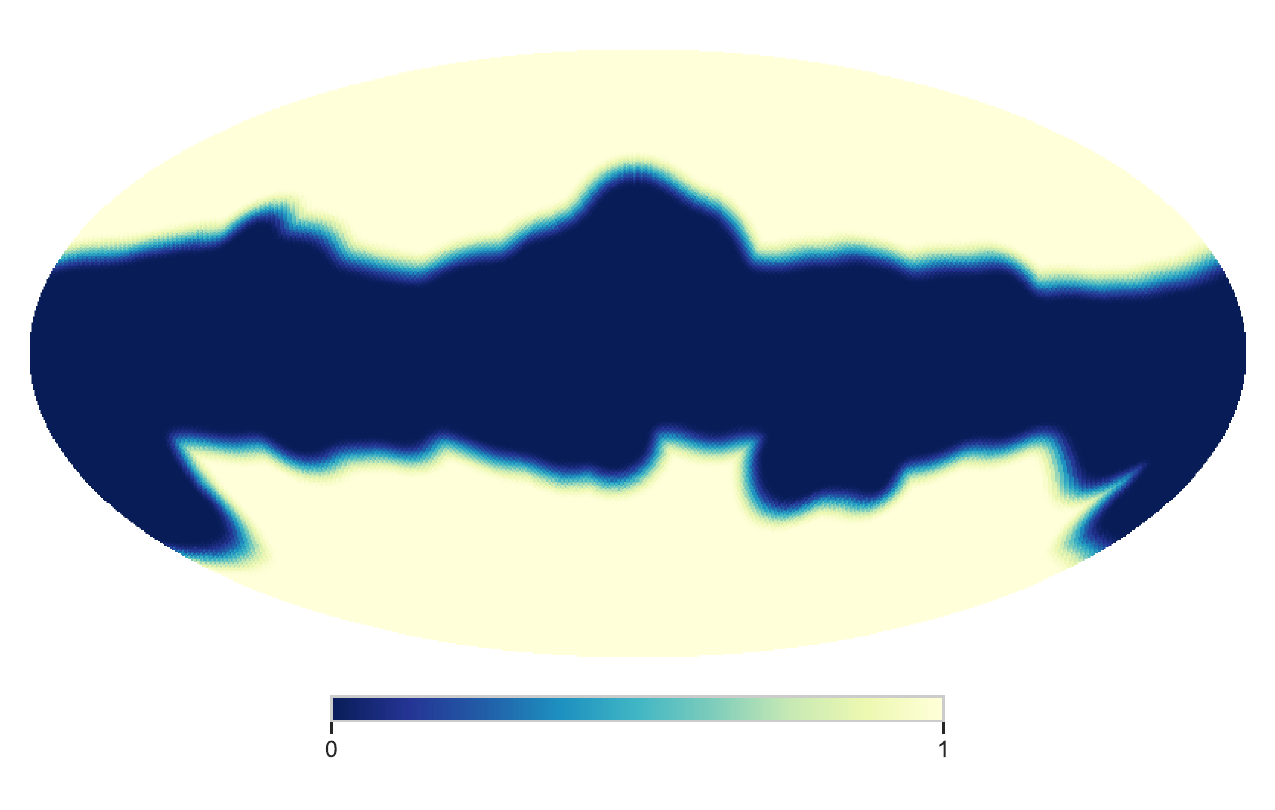}\label{fig:f2}}
  \caption{Left panel: Binary mask ($f_{\rm sky}=47\,\%$) used for the large-scale power spectrum estimation (QML) at $N_{\rm side}=16$ resolution. Right panel: \textit{Planck} Galactic-plane apodized mask ($f_{\rm sky}=51\,\%$) at $N_{\rm side}=64$ used for intermediate/small-scale pseudo-$C_{\ell}$s estimation  (see section \ref{sec:method} for details).}
  \label{fig:masks}
\end{figure}

\medskip
\noindent{\bf Sky masks.} We use two different masks for the two power spectrum estimators described above: a binary mask obtained from the {\sl LiteBIRD} polarization mask with sky fraction $f_{\rm sky}=49.5\,\%$ presented in section 5.2.4 of \cite{LiteBIRD:2022cnt} for the QML estimator and the \textit{Planck} Galactic plane mask appropriately apodized for the pseudo-$C_{\ell}$s (Fig.~\ref{fig:masks}).
We optimize the mask properties at both the spectrum and the parameter likelihood levels. First, we minimize both the variance of the $BB$ spectrum and the residuals of the simulations with respect to the input spectrum divided by the error on the mean at each multipole, while varying the threshold of the binary mask in the QML part and the apodization scale and sky fraction in the pseudo-$C_{\ell}$s one. We then verify that the mask configurations explored in the first step actually minimize the uncertainty and bias on the tensor-to-scalar ratio $r_{\rm vac}$ for a fiducial Starobinsky model (see also discussion below). We find the optimal configuration by downgrading the {\sl LiteBIRD} mask to $N_{\text{side}}=16$ and thresholding it at a value of 0.75, resulting in $f_{\rm sky}=47\,\%$ for the QML estimator, and by apodizing the \textit{Planck} $f_{\rm sky}=60\,\%$ mask at a $10^{\circ}$ scale with the $C^2$ method \cite{Grain:2009wq}, resulting in an effective $f_{\rm sky}=51\,\%$ for pseudo-$C_{\ell}$s.

\begin{figure}
    \centering
    \includegraphics[scale=0.5]{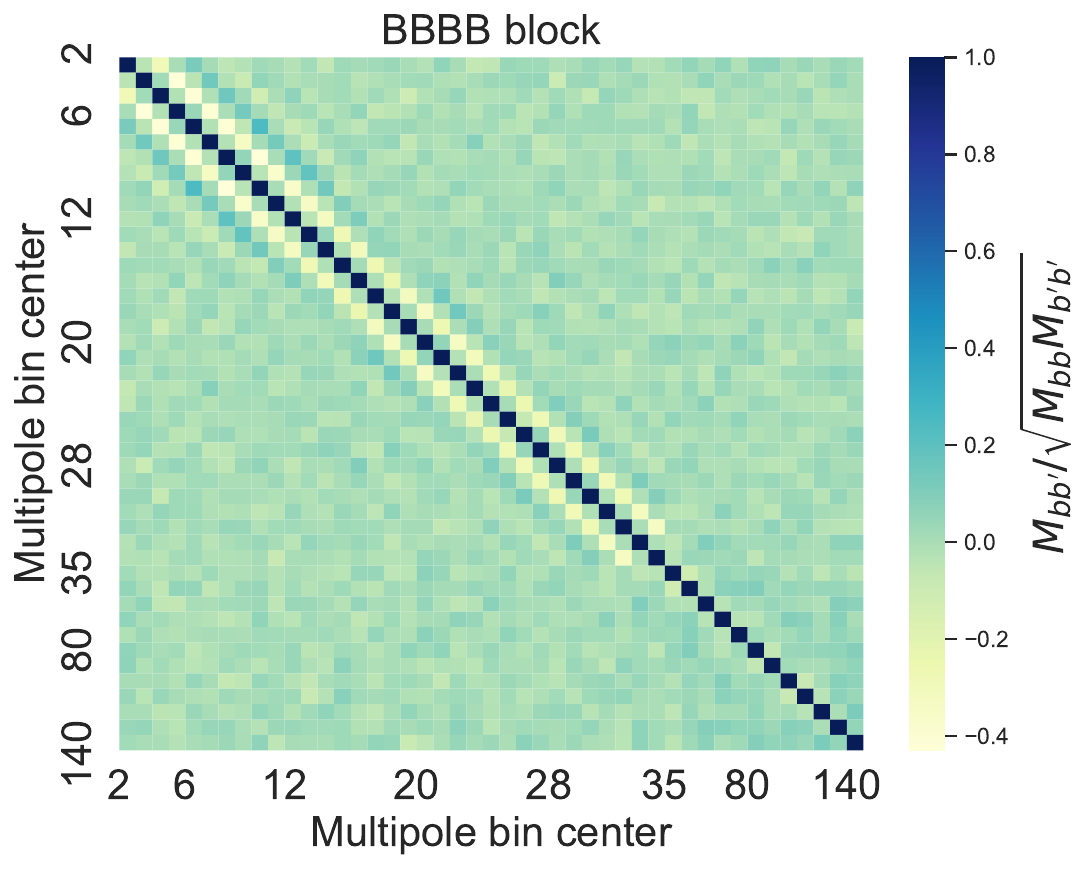}
    \caption{Correlation matrix for the BBBB covariance block.  Here multipoles up to $\ell=35$ are unbinned, while larger multipoles are binned with $\Delta\ell=10$.}
    \label{fig:covmat}
\end{figure}

\medskip
\noindent{\bf Likelihood.} We perform statistical inference on the inflationary model parameters using a likelihood based on the Hamimeche \& Lewis (hereafter H\&L) \cite{Hamimeche:2008ai} approximation. Specifically, we add an offset term, as prescribed in Ref.~\cite{Mangilli:2015xya}, to deal with negative values of the estimated power spectra on large scales. We also use the Sellentin \& Heavens (hereafter S\&H) \cite{Sellentin:2015waz} correction to the H\&L likelihood, accounting for the increased uncertainty in the parameters due to the finite number of simulations used in the estimation of the covariance matrix.
The block $BBBB$ of the $C_{\ell}$ covariance matrix estimated from simulations is shown in Fig.~\ref{fig:covmat}, with unbinned multipoles in $\ell<35$ and binned multipoles with $\Delta\ell=10$ for $\ell =  35 - 150$. Finally, we checked that the Monte Carlo noise due to the limited number of simulations used to estimate the covariance does not affect parameter inference: adopting the conditioning strategy suggested in Ref.~\cite{Beck:2022efr} leaves parameter estimates unchanged. All fits in this paper are performed by minimizing the negative logarithm of likelihood $-2\log\mathcal{L}$ by the multidimensional minimizer \texttt{iMinuit}\footnote{\url{iminuit.readthedocs.io}}.

\section{Disentangling vacuum and sourced origins: benefits of a full-sky mission}\label{sec:wrong_fit}

In this section, we highlight the benefits of a full-sky $B$-mode mission with access to the reionization bump scales in helping to understand the origin of primordial GW background. In the event of a detection of primordial $B$ modes, this feature could allow one to exclude quantum vacuum fluctuations of spacetime within the standard single-field slow-roll paradigm as the only source of the observed GW background, pointing instead towards production by matter sources, that is, the additional excited particle content active during inflation.
From this perspective, the axion-SU(2) model can provide a useful benchmark: it can source tensor modes that exceed by a factor $\mathcal{O}(5)$ the vacuum contribution at the largest scales, producing a pronounced feature in the reionization bump, while producing at the same time a recombination bump essentially indistinguishable from the standard single-field slow-roll prediction \cite{LiteBIRD:2022cnt} (Fig.~\ref{fig:ptep_updated}).

\begin{figure}
    \centering
    \includegraphics[scale=0.6]{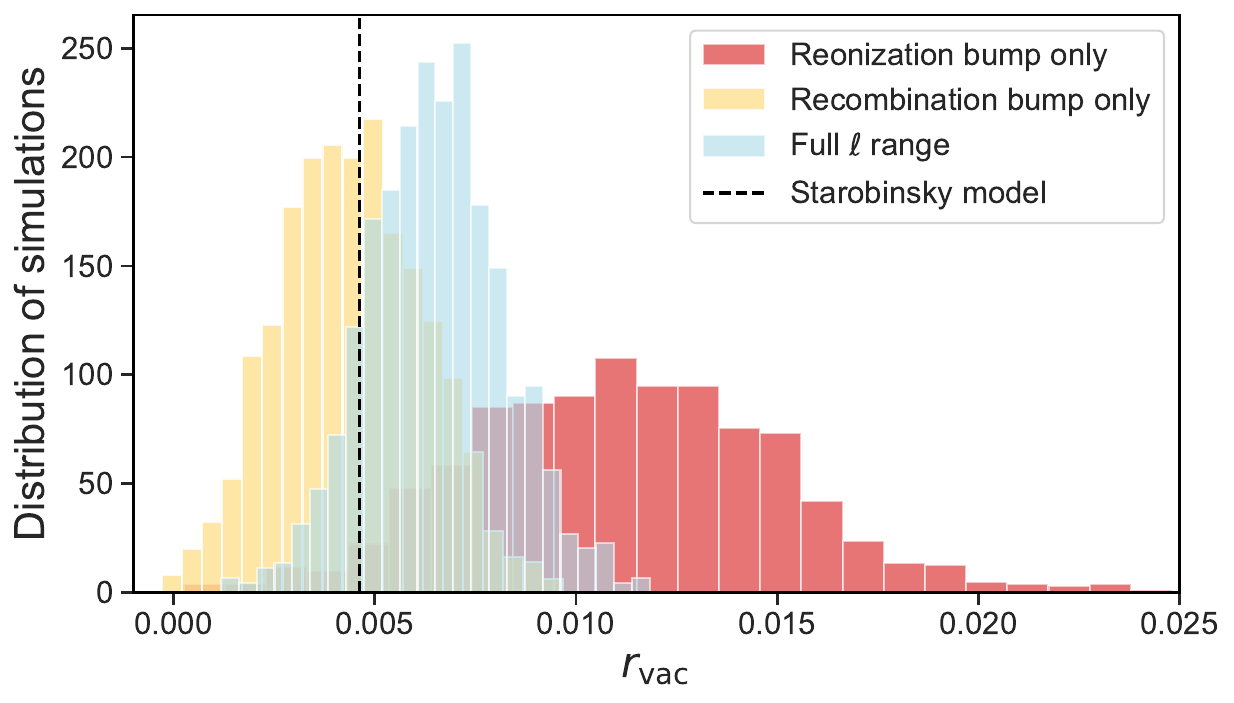}
    \caption{Histograms of the inferred vacuum tensor-to-scalar ratio $r_{\rm vac}$ values obtained from fitting an observed sky generated from the ``high reionization bump'' axion-SU(2) model (with parameters $r_*, \sigma, k_{\rm p}, r_{\rm vac}=[0.023, 1.1, 3.44\times10^{-4}\,\mathrm{Mpc}^{-1}, 0.00461]$) in three different ranges of multipoles: only the reionization bump range ($\ell=2-30$, in red), only the recombination bump range ($\ell=31-150$, in yellow) and the full range ($\ell=2-150$, light blue). We also show for reference the $r_{\rm vac}$ value for the Starobinsky model ($r_{\rm vac}=0.00461$).}
    \label{fig:pdf_wrong_fit}
\end{figure}

  \begin{table}
	\centering
	\begin{tabular}{lccc}
		%\hline \hline
	%
        &
		$\ell$ range
		&  $\Bar{r}_{\rm vac}$
		& $\sigma_{r_{\rm vac}}$
		\\
		\hline	%\hline
Full range & 2 -- 150 & \phantom{0}0.0065 & \phantom{0}0.0017 \\
Reionization bump & 2 -- 30 & 0.011 & 0.004 \\
 Recombination bump & 31 -- 150 & \phantom{0}0.0044 & \phantom{0}0.0018 \\
 		%\hline\hline
	\end{tabular}
	\caption{Mean values (indicated by $\Bar{r}_{\rm vac}$) and standard deviation of the distributions of the $r_{\rm vac}$ parameter in Fig.~\ref{fig:pdf_wrong_fit}, obtained by fitting a true sky generated from the ``high reionization bump'' SU(2) model in three different ranges of multipoles.}
	\label{table:wrong_fit}
\end{table}

Here we expand on the discussion in Ref.~\cite{LiteBIRD:2022cnt} in a more quantitative way. Suppose that the true sky was generated from the axion-SU(2) inflation. Then what if the observational data were fitted with a ``wrong'' model, in this case the standard single-field slow-roll one? To this end, we generate {\sl LiteBIRD} simulations for the ``high reionization bump'' parameter choice of the axion-SU(2) model (corresponding to the orange dot-dashed curve in Fig.~\ref{fig:ptep_updated}) using the procedure described in section \ref{sec:method}, and fit the $BB$ spectrum of each simulation for the tensor-to-scalar ratio parameter $r_{\rm vac}$ (as defined in section \ref{sec:theory}), assuming for simplicity that all other cosmological parameters are fixed by $TT$, $TE$ and $EE$. This is a good approximation, as the axion-SU(2) model provides negligible contributions to scalar perturbations, allowing us to break the degeneracy between the inflationary parameters and the $\Lambda$CDM parameters. The fit is performed three times for each simulation, taking three different ranges of multipoles of the $BB$ spectrum: only the reionization bump range ($\ell=2-30$), only the recombination bump range ($\ell=31-150$) and finally the full range of multipoles ($\ell=2-150$). The resulting histograms are shown in Fig.~\ref{fig:pdf_wrong_fit}, while we report in Table~\ref{table:wrong_fit} the mean value and standard deviation associated with the fit for each case.  As is evident, the fit recovers very different amplitudes for $r_{\rm vac}$ in the three cases, potentially leading to incorrect interpretations within the standard single-field slow-roll scenario. As was expected from our choice of the axion-SU(2) parameters, the data point towards the Starobinsky model if the survey does not have access to the reionization bump, even though the true sky features a significant contribution from matter sources.

 We note that if we also fit, in addition to $r_{\rm vac}$, the underlying ``high reionization bump'' SU(2) simulations for the tensor spectral index $n_{\rm t}$, we expect $n_{\rm t}$ to be partially degenerate with $r_*$, $\sigma$ and $k_{\rm p}$. In this case, the recovered $n_{\rm t}$ would be very red, violating the standard single-field slow-roll consistency relation and pointing towards the presence of more than one field during inflation. Checking the Gaussianity of the signal would then be essential in discriminating between axion-SU(2) and other models (see section \ref{sec:intro} for more details).

\section{Constraining axion-SU(2) parameters with  LiteBIRD}\label{sec:constraints}
In this section, we explore the possibility of constraining the parameters of the axion-SU(2) model using the {\sl LiteBIRD} satellite. We infer parameters using a frequentist approach based on the Monte Carlo simulations constructed in section \ref{sec:method}, and account for the presence of physical boundaries on the parameters following the Feldman-Cousins prescription \cite{Feldman:1997qc}.
We start in section \ref{sec:feldman} by describing the Feldman-Cousins approach and applying it to our case study, and in section \ref{sec:robustness} we explore the robustness of this construction. 

\subsection{Feldman-Cousins approach}\label{sec:feldman}

Frequentist confidence intervals for the parameters of interest are built using the classical Neyman's construction \cite{Neyman}.   
However, if the parameter has a physical boundary and its estimate is close to this boundary, Neyman's construction must be corrected as indicated by Feldman \& Cousins \cite{Feldman:1997qc} (hereafter FC). This approach is a standard staple in particle physics data analysis \cite{ATLAS:2013mma}.  Recently there has been renewed interest in using it for cosmology, being applied, for example, to \textit{Planck} data \cite{planck_profile}, SPIDER $B$-mode data \cite{SPIDER:2021ncy}, early dark energy models \cite{Herold:2021ksg}, the tensor-to-scalar ratio from the \textit{Planck} and BICEP/Keck data \cite{Campeti:2022vom}, and the search for signatures of axion-U(1) inflation in the current data \cite{Campeti:2022acx}. This technique is not plagued by prior volume effects that can arise during the marginalization procedure in the Monte Carlo Markov Chain (MCMC) approach, when performing inference on degenerate/unconstrained parameters \cite{Hamann:2007pi, Herold:2021ksg, Campeti:2022vom}. The Feldman-Cousins approach is therefore an ideal choice for our case study, since we expect the axion-SU(2) model parameters to be degenerate and hard to constrain simultaneously.

The FC recipe for our parameter of interest, $r_*$, is the following:
\begin{enumerate}
    \item For each of the  input values in the physically-allowed range of interest, $r_{*}^{\rm in}$, we generate a sufficient number of simulations.
    \item We fit a model to each simulation, obtaining values of $r_{*}^{\rm mle}$ that maximize the likelihood given the input simulation. These can be used to build a histogram of $r_{*}^{\rm mle}$ values for each input $r_{*}^{\rm in}$, which is then used in Step 3.
    \item We determine, for each $r_{*}^{\rm in}$, the interval of $r_{*}^{\rm mle}$ values that includes 95\,\% of the simulations (for a 95\,\% C.L. confidence interval) with the highest \textit{likelihood ratio}\footnote{The likelihood ratio is actually a \textit{profile likelihood ratio} \cite{cranmer, ParticleDataGroup:2020ssz}, where one uses the maximum likelihood estimates of the nuisance parameters in Eq.~\ref{eq:lkl_ratio}.}, defined as
\begin{equation}\label{eq:lkl_ratio}
    R(r_{*}^{\rm mle}, r_{*}^{\rm in}) = \frac{\mathcal{L}(r_{*}^{\rm mle}|r_{*}^{\rm in})}{\mathcal{L}(r_{*}^{\rm mle}|r_{*}^{\rm best})},
\end{equation}
where $r_{*}^{\rm best}$  is the value of $r_{*}^{\rm in}$ that maximizes the likelihood $\mathcal{L}(r_{*}^{\rm mle}|r_{*}^{\rm in})$. This determines a horizontal interval at each $r_{*}^{\rm in}$ in the $r_{*}^{\rm in}$ versus $r_{*}^{\rm mle}$ plane: the union of all these intervals creates the \textit{confidence belt}.
\item Finally, we intersect the confidence belt by drawing a vertical line at the observed value of $r_{*}^{\rm mle}$, which we call $r_{*}^{\rm obs}$. We can then read off the $r_{*}^{\rm in}$ axis an upper (lower) limit if the confidence belt is intersected only in its upper (lower) part or a two-sided confidence interval if there are two intersection points. 
\end{enumerate}
In our specific case, we take 200 input values of $r_{*}^{\rm in}$ linearly spaced in the range $[0,\,0.2]$ and generate for each value 1000 {\sl LiteBIRD} simulations as described in section \ref{sec:method}, keeping fixed $k_{\rm p}=3.44\times 10^{-4}\,\mathrm{Mpc}^{-1}$, $\sigma=1.1$ and $r_{\rm vac}=0.00461$. We use $r_{*}^{\rm obs}=0.015$ as the observed value, obtained as the best fit to the simulation generated from the ``high reionization bump'' parameters (hereafter ``ground truth'' values) with $r_{*}^{\rm in}=0.023$. 

\begin{figure}
    \centering
    \includegraphics[scale=0.7]{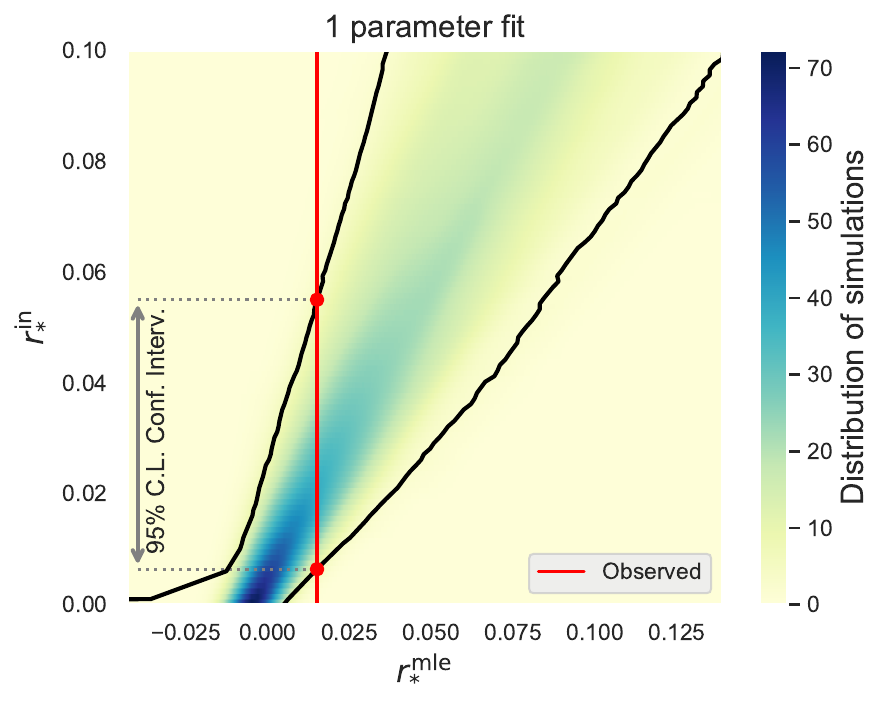}
    \caption{FC construction for the 1-parameter fit (solid black lines), with $\sigma$ and $k_{\rm p}$ fixed to their ground truth. {\sl LiteBIRD} will be able to obtain a two-sided 95\,\% C.L. confidence interval $r_*=0.015^{+0.04}_{-0.008}$, if the observed sky has been generated from the ``high reionization bump'' model. The observed value $r_{*}^{\rm obs} = 0.015$ is indicated as a vertical solid red line, and its intersection with the confidence belt with two red dots. The color bar shows the distribution of $r_{*}^{\rm mle}$ obtained by fitting the simulations as a function of the input $r_{*}^{\rm in}$. See section \ref{sec:feldman} for details.}
    \label{fig:beauty_full_1param}
\end{figure}

In Fig.~\ref{fig:beauty_full_1param}, we show the FC construction (solid black lines) in which we fit the simulations only for $r_{*}$ (hereafter 1 parameter fit), fixing $\sigma$ and $k_{\rm p}$ to their ground truth in the template (Eq.~\ref{eq:template}). In this case, we observe that {\sl LiteBIRD} will be able to obtain a two-sided 95\,\% C.L. confidence interval of $r_*=0.015^{+0.04}_{-0.008}$, if the observed sky has been generated from the ``high reionization bump'' model. 
In the next section,  we will discuss the robustness of these constraints against a mismatch between the assumed values of $\sigma$ and $k_{\rm p}$ in the fit and the ground truth, as well as the degradation of the constraint in $r_*$ when we also treat $\sigma$ and $k_{\rm p}$ as free parameters in the fit.

\subsection{Robustness of the constraints}\label{sec:robustness}

In this section, we will explore how constraints on $r_{*}$ from the FC construction can change if we adopt ``wrong'' assumptions about $\sigma$ and $k_{\rm p}$ in the 1-parameter fit. We also assess the degradation of the constraints on $r_*$ when $\sigma$ and $k_{\rm p}$ are additional free parameters in the fit, exploiting the full power of the FC construction: the parameter inference performed with this technique is free from prior volume effects due to degeneracies/unconstrained parameters in the Bayesian MCMC \cite{Hamann:2007pi, Herold:2021ksg, Campeti:2022vom}.

\begin{figure}
    \centering
    \includegraphics[scale=0.7]{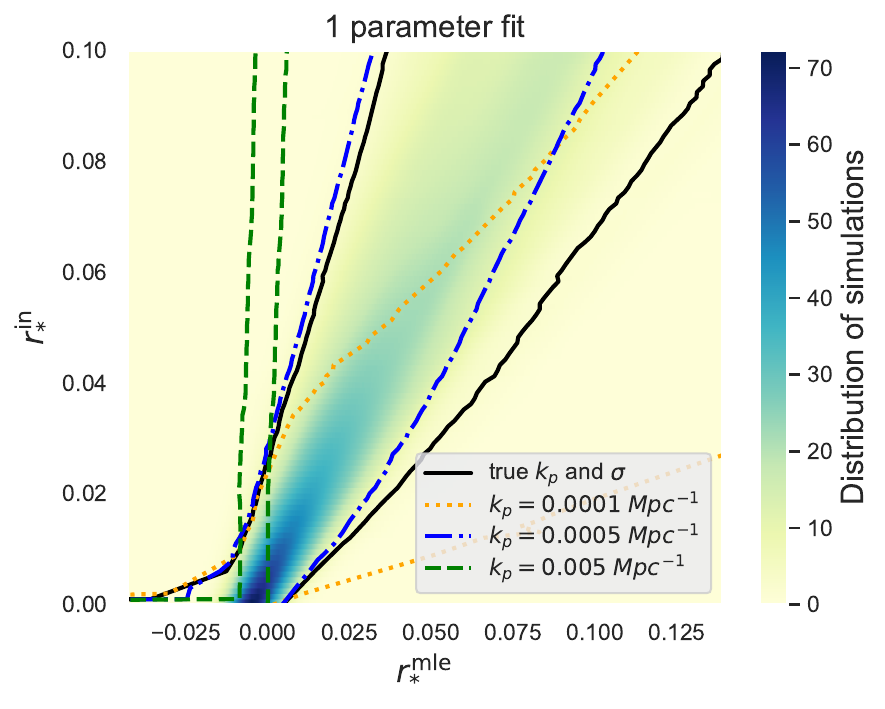}
    \caption{FC construction for the 1-parameter fit with $k_{\rm p}$ fixed to values different from its ground truth, i.e. $k_{\rm p}=[10^{-4},\, 5\times 10^{-4},\, 5\times 10^{-3}]\,\mathrm{Mpc}^{-1}$ (dotted orange, dot-dashed blue and dashed green, respecitvely), compared to the ground truth confidence belt (solid black, same as in Fig.~\ref{fig:beauty_full_1param}). We assume the ground truth for $\sigma$. The color bar shows the distribution of $r_{*}^{\rm mle}$ obtained by fitting the simulations (assuming the ground truth for $k_p$ and $\sigma$) as a function of the input $r_{*}^{\rm in}$. See section \ref{sec:robustness} for details.}
    \label{fig:beauty_full}
\end{figure}

We start by examining the FC confidence belt built from the same simulations as in section \ref{sec:feldman} and fitting each one for $r_{*}$; however, in this case, we consider three different values of the peak scale $k_{\rm p}=[10^{-4},\, 5\times 10^{-4},\, 5\times 10^{-3}]\,\mathrm{Mpc}^{-1}$ as shown in Fig.~\ref{fig:beauty_full}, instead of fixing it to its ground truth, $k_{\rm p}=3.4\times 10^{-4}\, \mathrm{Mpc}^{-1}$. We still fix $\sigma$ to its ground truth; similar conclusions can be drawn by varying $\sigma$ instead of $k_{\rm p}$.
The behavior of the confidence belts in Fig.~\ref{fig:beauty_full} can be easily interpreted. The larger $r_{*}^{\rm in}$ is, the larger the power in the underlying simulations at the reionization bump becomes. However, if we fix $k_{\rm p}=5\times 10^{-3} \,\mathrm{Mpc}^{-1}$ in the fit (green dashed), the template model can add power only at recombination bump scales ($k\sim 5\times 10^{-3}\, \mathrm{Mpc}^{-1}$): this results in a preference for small values $r_{*}^{\rm mle}$ around zero, since the power in simulations at recombination bump scales is already saturated by $r_{\rm vac}$.  
Similarly, for the $k_{\rm p}=5\times 10^{-4}\, \mathrm{Mpc}^{-1}$ case (dot-dashed blue), the confidence belt is slightly tighter compared to the initial one (solid black) as less power is allowed at smaller scales compared to reionization bump scales in the simulations, while the opposite is true for the $k_{\rm p}=10^{-4} \,\mathrm{Mpc}^{-1}$ case (dotted orange).  
The best choice of $k_{\rm p}$ can be found by comparing the value of the likelihood for each model considered.

\begin{figure}
    \centering
    \includegraphics[scale=0.7]{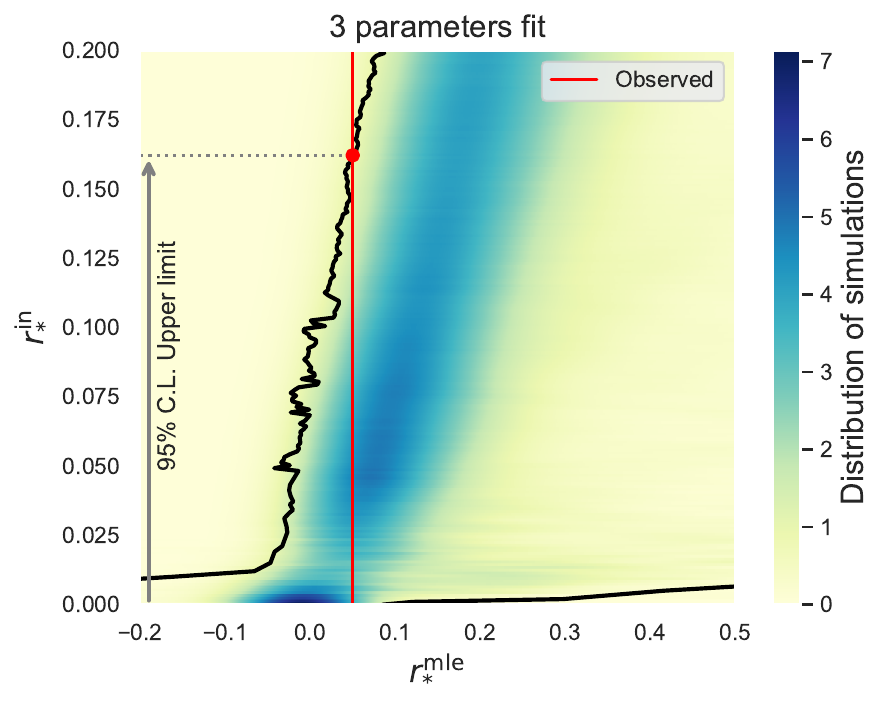}
    \caption{FC construction for the 3-parameter fit (solid black lines) with free parameters $r_*$, $\sigma$ and $k_{\rm p}$. {\sl LiteBIRD} will be able to obtain a 95\,\% C.L. upper limit $r_{*}\leq 0.16$, if the observed sky has been generated from the ``high reionization bump'' model. The observed value $r_{*}^{\rm obs} = 0.05$ is indicated as a vertical solid red line. The color bar shows the distribution of $r_{*}^{\rm mle}$ obtained by fitting the simulations as a function of the input $r_{*}^{\rm in}$. See section \ref{sec:robustness} for details.}
    \label{fig:beauty_full_3param}
\end{figure}

Now, we check how much the constraints presented in section \ref{sec:feldman} are degraded when also $\sigma$ and $k_{\rm p}$ values re allowed to vary in the fit. In Fig.~\ref{fig:beauty_full_3param}, we show the confidence belt for the 3-parameter fit ($r_*, \sigma, k_{\rm p}$). In this case, {\sl LiteBIRD} will be able to put a 95\,\% C.L. upper limit on $r_{*}\leq 0.16$, if the observed sky has been generated from the ``high reionization bump'' model. This significant degrading of the constraining power is mainly due to the degeneracy among the three parameters of the model given in Eq.~\ref{eq:template}.
We also compare in Table~\ref{table:tab1} the 95\,\% C.L. confidence interval or upper limit on $r_*$ obtained from the FC construction in the 1-, 2- and 3-parameter fits. We report the observed value $r_{*}^{\rm obs}$ and the corresponding error bars or upper limits. For the 1-parameter fit, we fix $\sigma$ and $k_{\rm p}$ to their ground truth values, while in the 2-parameter case we additionally fit for $\sigma$ while $k_{\rm p}$ is fixed to the ground truth value.

 \begin{table*}
	\centering\tabcolsep=0.7cm
 \renewcommand{\arraystretch}{1.3} % General space between rows (1 standard)
	\begin{tabular}{ll}
	 
		& Constraint on $r_{*}$\\
		\hline
 1-parameter fit ($r_*$)&  $r_*=0.015^{+0.04}_{-0.008}$ \\
2-parameter fit ($r_*, \sigma$) & $r_*=0.033^{+0.06}_{-0.028}$ \\ 
3-parameter fit ($r_*, \sigma, k_{\rm p}$) & $r_{*}^{\rm obs} = 0.05$; $r_{*}\leq 0.16$\\
	\end{tabular}
	\caption{95\,\% C.L. confidence intervals or uppper limits on $r_*$ obtained from the FC construction in the 1-, 2- and 3-parameter fits. We assume as the underlying model the ``high reionization bump'' model (section \ref{sec:theory}). See section \ref{sec:robustness} for more details. 
}
	\label{table:tab1}
\end{table*}

\section{Parity-violating $TB$ and $EB$ correlations}\label{sec:TBEB}

In this section, we investigate another unique signature of the axion-SU(2) inflation model when trying to disentangle it from standard single-field slow-roll models: $TB$ and $EB$ parity-violating correlations in the CMB produced by chiral gravitational waves \cite{Lue:1998mq, Shiraishi:2016yun, Saito:2007kt, Gluscevic_2010, Gerbino:2016mqb}. 
We finally assess the full power of {\sl LiteBIRD} in discriminating between two mechanisms for enhanced gravitational wave production, using all CMB spectra ($TT$, $EE$, $BB$, $TE$, $TB$ and $EB$) in the covariance matrix. 

Concerning $TB$ and $EB$ spectra, we improve and update the {\sl LiteBIRD} forecast presented in \cite{Thorne:2017jft} in a number of aspects. First, we relax the assumption used in \cite{Thorne:2017jft} of the same CMB upper bound on $r_{\rm vac}$ applying to all scales. Instead, we take into account that at reionization bump scales this bound becomes considerably relaxed compared to the one at at $k=0.05\,\mathrm{Mpc}^{-1}$ \cite{Tristram:2020wbi, Tristram:2021tvh, deBelsunce:2022yll}. This allows, in principle, for a larger signal from the axion-SU(2) model at these scales. Second, we update the forecast with new bounds on the SU(2) parameter space from backreaction \cite{Ishiwata:2021yne} and the current upper limits from the \textit{Planck} and BICEP data \cite{Tristram:2021tvh, Hamann:2022apw, Campeti:2022vom, BICEP:2021xfz}. Third, we show results for the full covariance matrix from realistic {\sl LiteBIRD} simulations, instead of the simplified Fisher approach of \cite{Thorne:2017jft}.
Fourth, since we found that the signal-to-noise ratio defined in \cite{Thorne:2017jft} is strongly dependent on the fiducial model adopted in the covariance matrix, we use as a figure of merit $\Delta\chi^2$ between the standard single-field slow-roll fiducial  (specifically, the Starobinsky model) and axion-SU(2) models, that is 
\begin{equation}\label{eq:chi2}
    \Delta\chi^2 = \left(\mathbf{C}_{\ell}^{\rm SU(2)} - \mathbf{C}_{\ell}^{\rm fid} \right)^{\rm T} \mathbf{M}^{-1} \left(\mathbf{C}_{\ell}^{\rm SU(2)} - \mathbf{C}_{\ell}^{\rm fid} \right), 
\end{equation}
where $\mathbf{M}$ is the $C_{\ell}$ covariance matrix, $\mathbf{C}_{\ell}^{\rm fid}$ is a vector containing the fiducial angular power spectra and similarly $\mathbf{C}_{\ell}^{\rm SU(2)}$ contains the theoretical spectra for the axion-SU(2) model.

\begin{figure}
\begin{center}
\includegraphics[width=.49\textwidth]{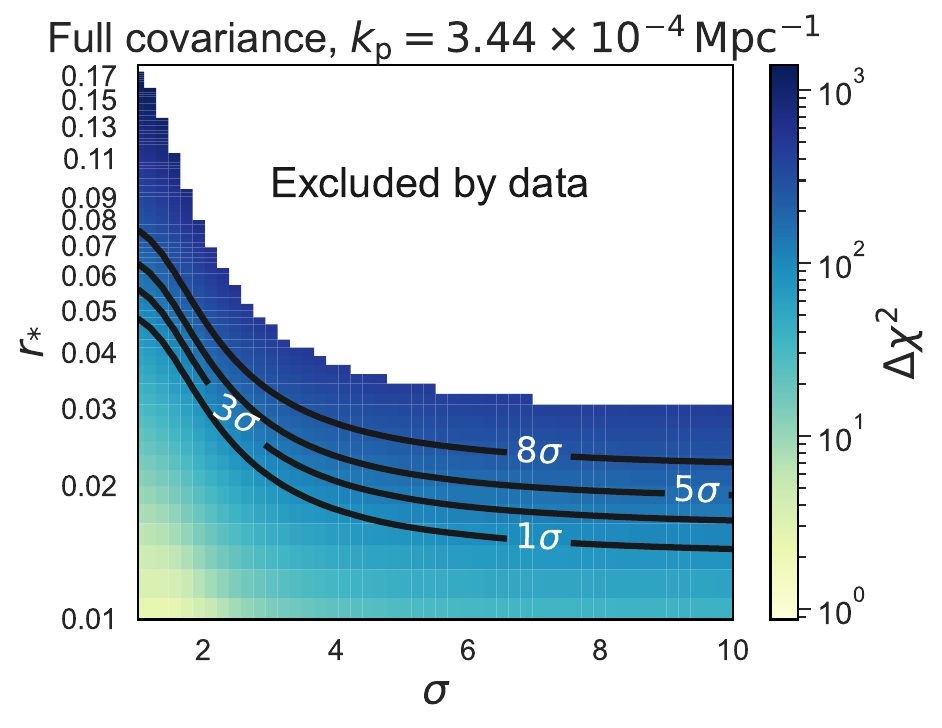}
\includegraphics[width=.49\textwidth]{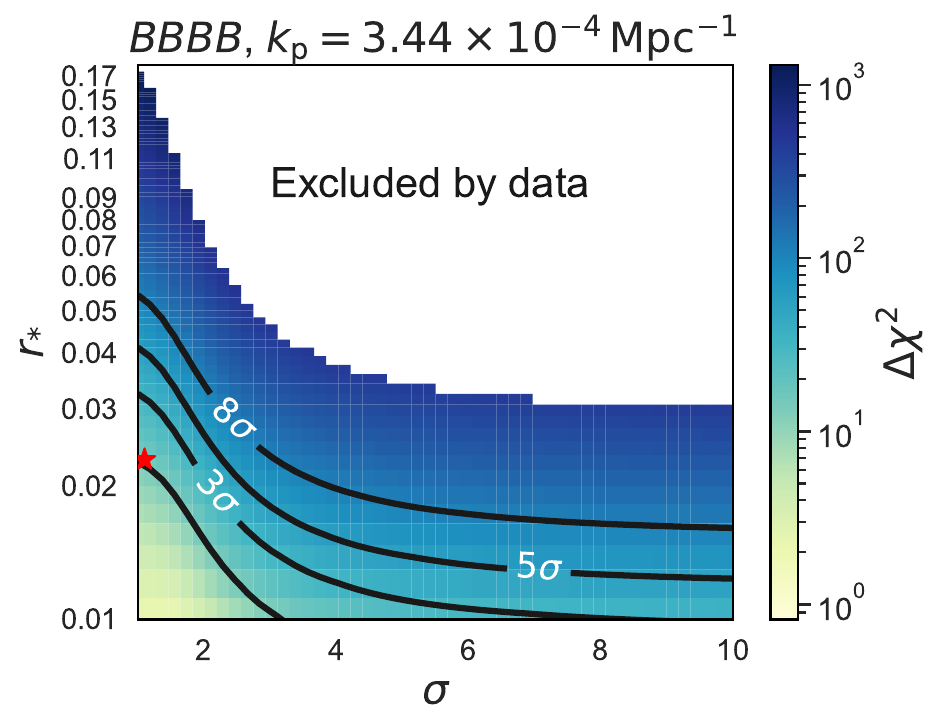}
\includegraphics[width=.49\textwidth]{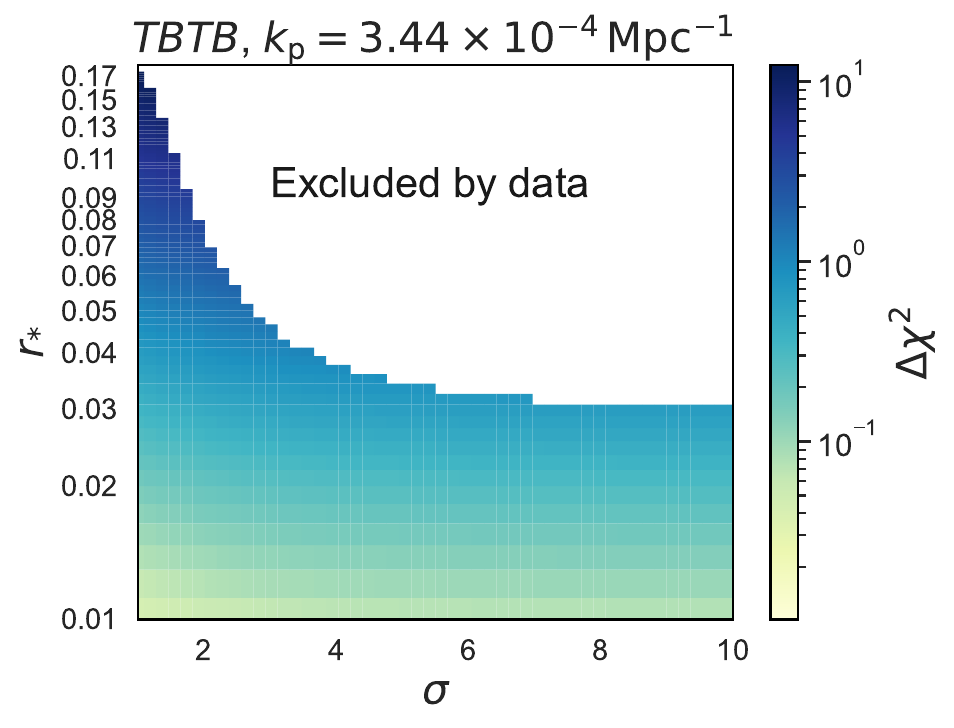}
\includegraphics[width=.49\textwidth]{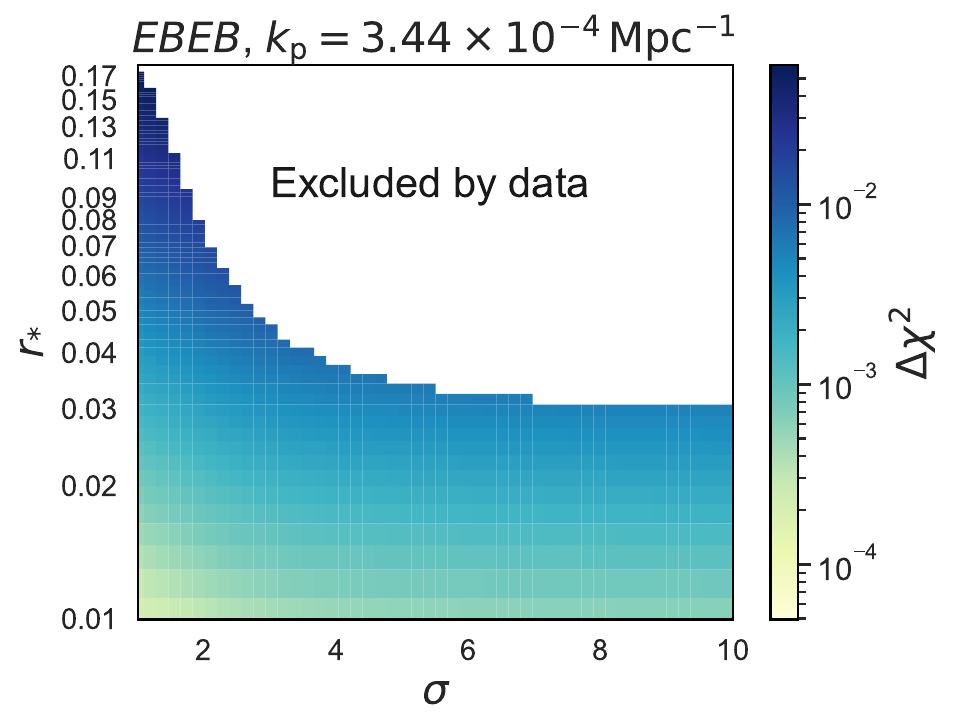}
\end{center}
    \caption{Comparison of $\Delta\chi^2$ (Eq.~\ref{eq:chi2}) between the axion-SU(2) and standard single-field slow-roll Starobinsky models from the full covariance matrix (top left panel), $BBBB$ (top right), $TBTB$ (bottom left) and $EBEB$ (bottom right) covariance blocks for {\sl LiteBIRD}, as a function of $r_*$ and $\sigma$ given $k_{\rm p}=3.44\times\,\mathrm{Mpc}^{-1}$. The contours show $1\,\sigma$, $3\,\sigma$, $5\,\sigma$ and $8\,\sigma$ significance (note that this $\sigma$, although named the same, is not the parameter of the axion-SU(2) model, which is instead plotted on the $x$-axis).  The white area is excluded by current upper limits on the tensor-to-scalar ratio. The red $\star$ symbol in the upper right panel indicates the ``high reionization bump'' model. Note that the color scale changes in every panel.}
    \label{fig:delta_chi_onescale}
\end{figure}

\begin{figure}
\begin{center}
\includegraphics[width=.49\textwidth]{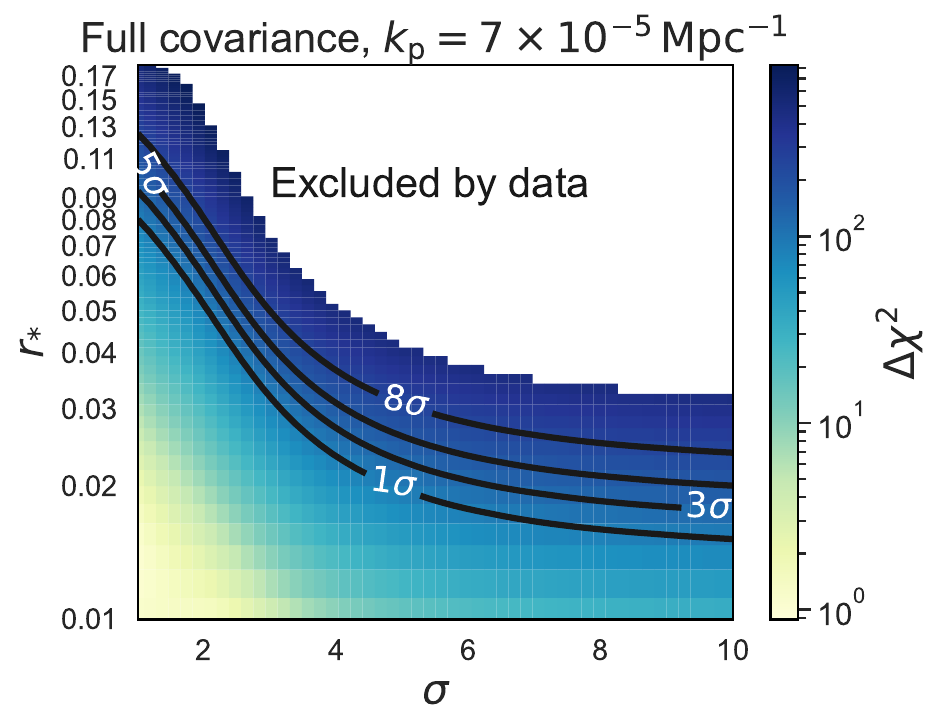}
\includegraphics[width=.49\textwidth]{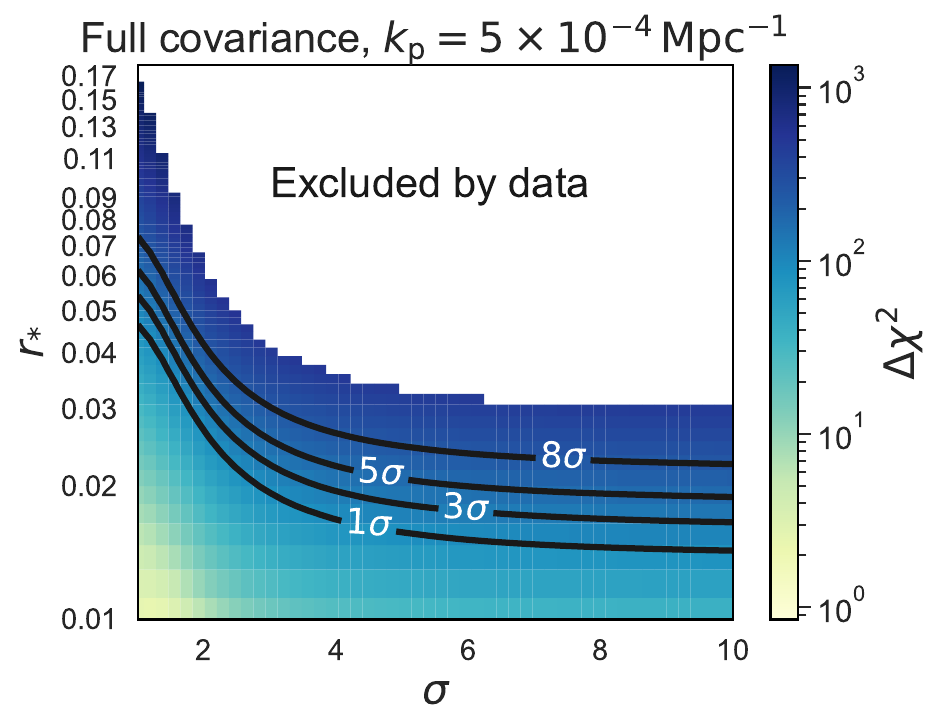}
\includegraphics[width=.49\textwidth]{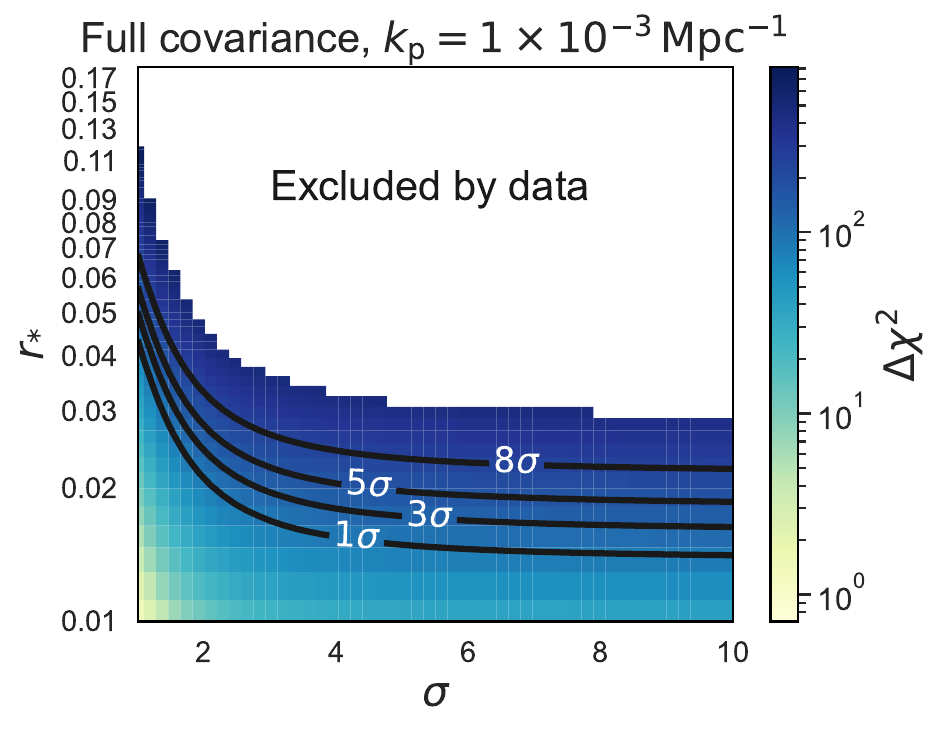}
\includegraphics[width=.49\textwidth]{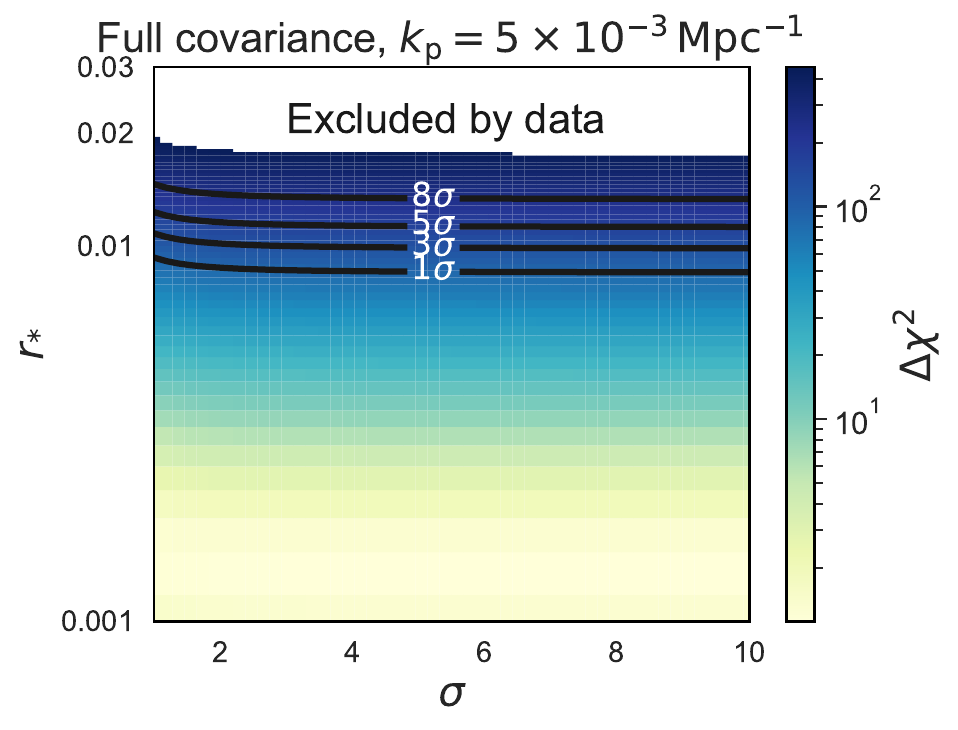}
\end{center}
    \caption{Same as Fig.~\ref{fig:delta_chi_onescale} but each panel has a different $k_{\rm p}$ and we use the full covariance matrix for all panels. Here we assume $r_{*}=5r_{\rm vac}$ in the top two and bottom left panels, while the bottom right panel assumes $r_{*}=r_{\rm vac}$, following Ref.~\cite{Ishiwata:2021yne}. Note that the color scale changes in every panel and $y$-axis range changes in the bottom right panel.}
    \label{fig:delta_chi_scales}
\end{figure}

We quantify in Figs.~\ref{fig:delta_chi_onescale} and \ref{fig:delta_chi_scales} the discriminating power of {\sl LiteBIRD} between the axion-SU(2) and standard single-field slow-roll Starobinsky models ($r_{\rm vac}=0.00461$) via $\Delta\chi^2$ values as a function of $r_*$ and $\sigma$ given a fixed scale $k_{\rm p}$. For each point in this parameter space, we compute the quantity $\Delta\chi^2$ in Eq.~\ref{eq:chi2}: larger positive values of $\Delta\chi^2$ indicate an increased ability of LiteBIRD to discriminate axion-SU(2) from the fiducial model. The associated $p$-value, i.e. the probability that the $\chi^2$ statistic should exceed a particular value $\Delta\chi^2$ by chance, given that the null hypothesis (i.e. the fiducial model) is correct, can then be computed as \cite{ParticleDataGroup:2020ssz} 
\begin{equation}\label{eq:p-value}
    p(\Delta\chi^2, n) =\int_{\Delta\chi^2}^{+\infty} \chi^2_n(x) dx,
\end{equation}
where $\chi^2_n(x)$ is the distribution for $n$ degrees of freedom (hereafter d.o.f.). In our analysis, the number of d.o.f. will be equal to the total number of multipole bins in all spectra considered\footnote{For this analysis we bin angular power spectra with equal width $\Delta\ell=10$, resulting in 14 bins for each spectrum. We additionally checked that the distribution of ${\Delta\chi^2_{{\rm null},\,i}=\left(\mathbf{C}_{\ell}^{{\rm sim},\, i} - \mathbf{C}_{\ell}^{\rm fid} \right)^{\rm T} \mathbf{M}^{-1} \left(\mathbf{C}_{\ell}^{{\rm sim},\, i} - \mathbf{C}_{\ell}^{\rm fid} \right)}$ for fiducial Starobinsky model simulations (i.e., the null hypothesis), where $\mathbf{C}_{\ell}^{{\rm sim},\, i}$ is the spectrum estimated from the $i$th simulation, follows a $\chi^2$ distribution with the expected number of d.o.f.}. The $p$-value can be converted into an equivalent significance $Z$, defined such that a Gaussian distributed random variable, fluctuating $Z$ standard deviations above its mean, has an upper-tail probability equal to $p$, that is, $Z=\Phi^{-1}(1-p)$, where $\Phi$ is the cumulative distribution of the standard Gaussian and $\Phi^{-1}$ its inverse quantile \cite{ParticleDataGroup:2020ssz}. For instance, $Z=5$ (i.e. a $5\,\sigma$ detection) corresponds to $p=2.87\times 10^{-7}$.  
We use this relation to draw reference contours of significance of $1\,\sigma$, $3\,\sigma$, $5\,\sigma$ and $8\,\sigma$ in Figs.~\ref{fig:delta_chi_onescale} and \ref{fig:delta_chi_scales}. Note that this $\sigma$, although named the same, is not the parameter $\sigma$ of the axion-SU(2) model, which is instead plotted on the $x$-axis of these figures.

\begin{figure}
\begin{center}
\includegraphics[width=1\textwidth]{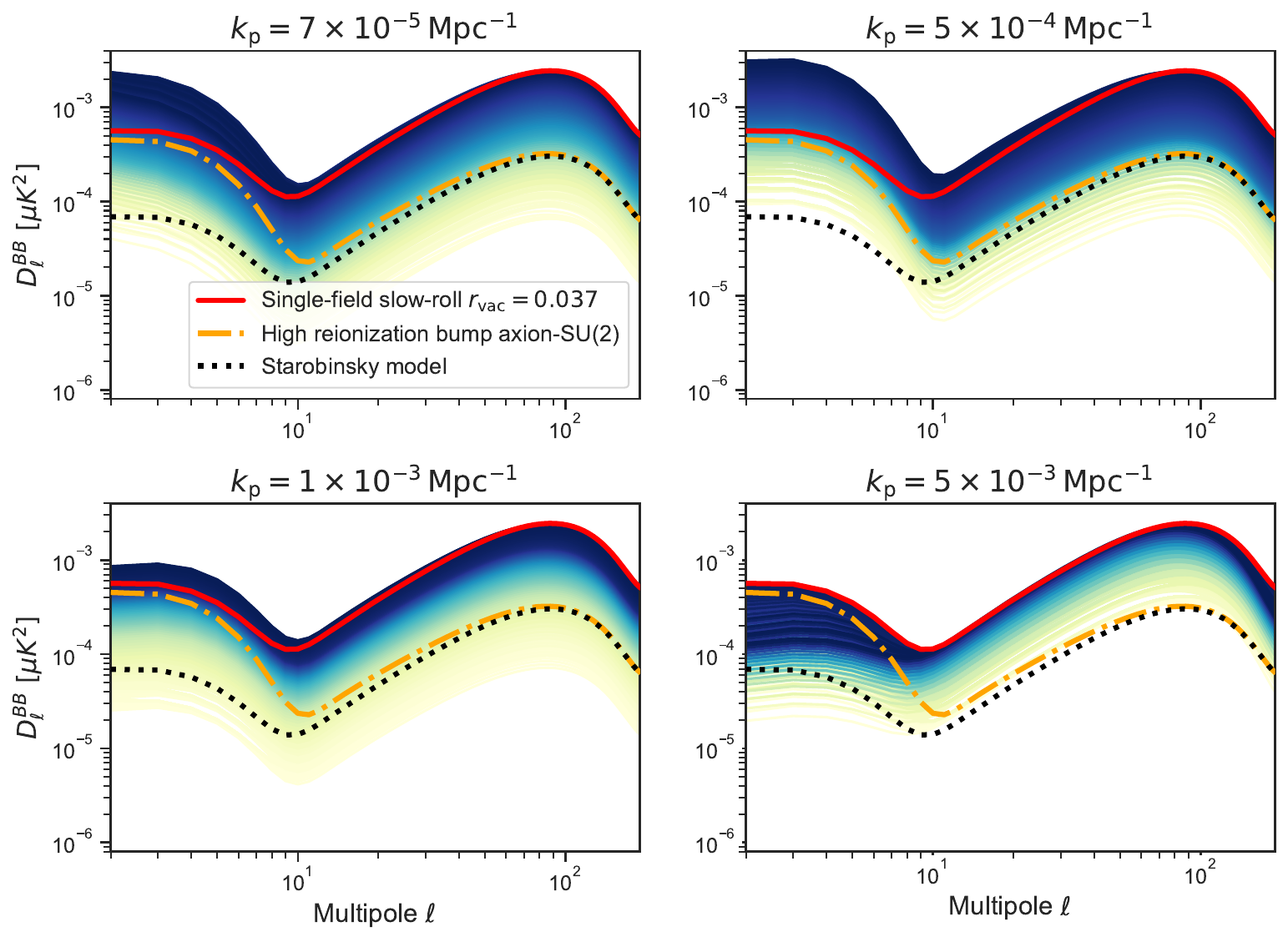}
\end{center}
    \caption{$BB$ power spectra for each of the parameter sets used in each panel of Fig.~\ref{fig:delta_chi_scales}. They correspond to spectra computed on a grid of 100 linearly spaced values of $r_*$ in the range $0.001-0.36$ and 50 values of $\sigma$ in the range $1-10$ at each fixed $k_{\rm p}$ value, excluding spectra not compatible with theoretical consistency arguments and observational bounds (see section \ref{sec:theory} for details). Darker color lines correspond to larger $r_*$ and $\sigma$ in this range.  We also show for reference the $BB$ spectrum for standard single-field slow-roll inflation obtained for $r_{\rm vac}=0.037$ (solid red), saturating the current upper at scales $0.05\,\mathrm{Mpc}^{-1}$, the ``high reionization bump'' axion-SU(2) model (dot-dashed orange, see section\ref{sec:theory}) and the Starobinsky model (dotted black).}
    \label{fig:plot_raphael}
\end{figure}

\begin{figure}
\begin{center}
\includegraphics[width=1\textwidth]{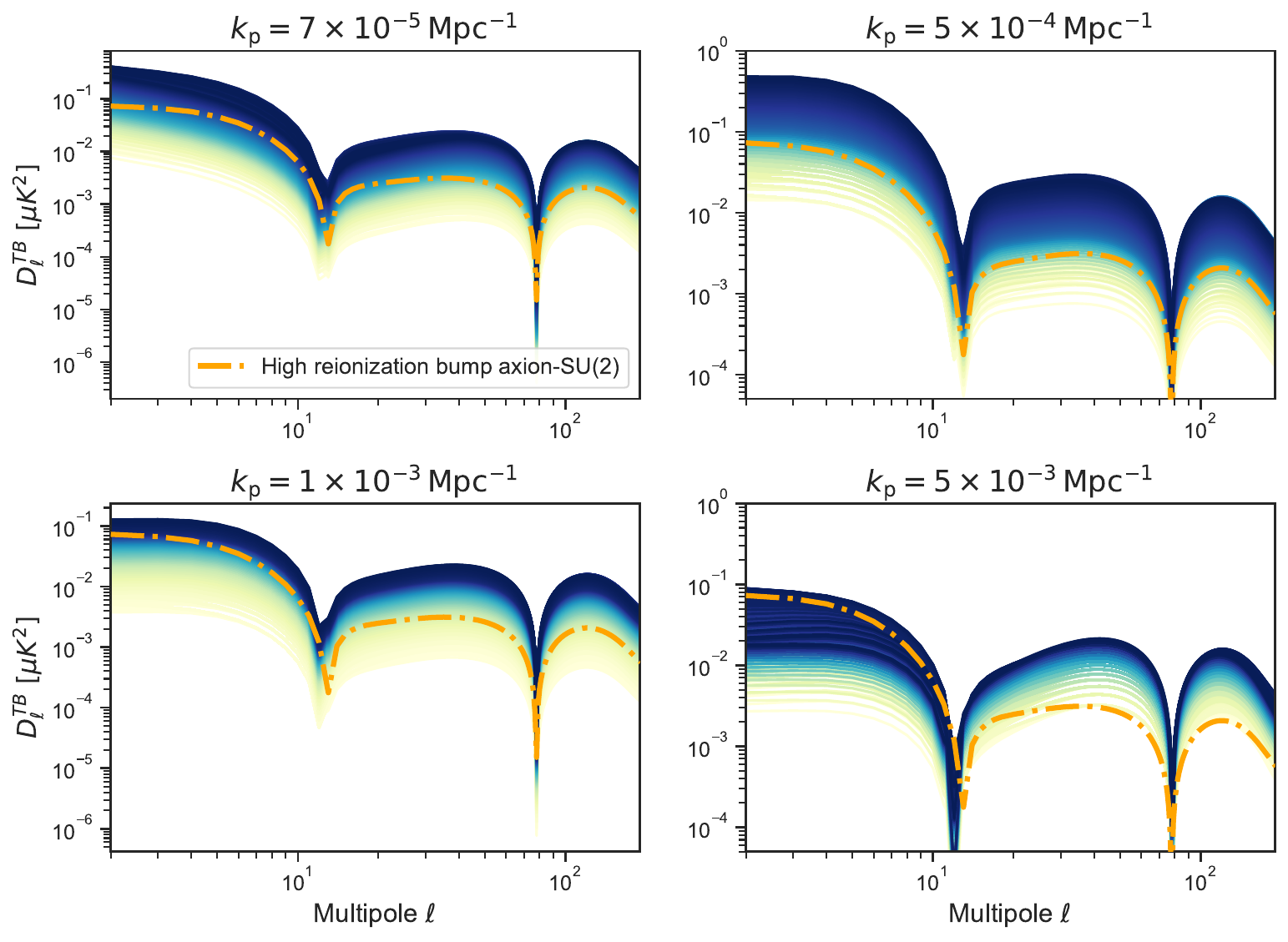}
\end{center}
    \caption{Same as Fig.~\ref{fig:plot_raphael} but for $|TB|$ power spectra. We also show for reference the $|TB|$ spectrum for the ``high reionization bump'' axion-SU(2) model (dot-dashed orange, see section~\ref{sec:theory}). }
    \label{fig:plot_raphael_TB}
\end{figure}

\begin{figure}
\begin{center}
\includegraphics[width=1\textwidth]{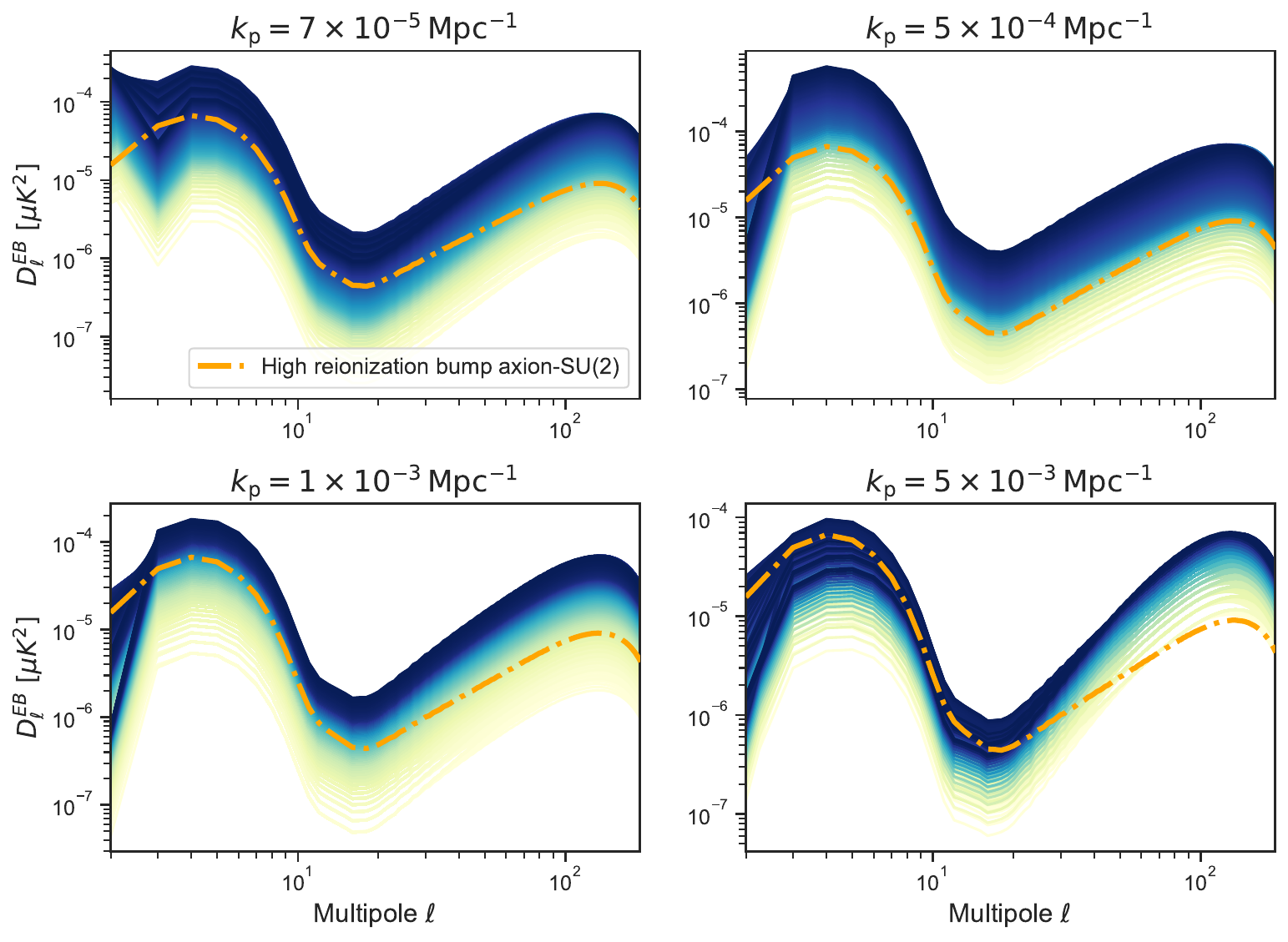}
\end{center}
    \caption{Same as Fig.~\ref{fig:plot_raphael_TB} but for $|EB|$ power spectra. 
    }
    \label{fig:plot_raphael_EB}
\end{figure}

In Fig.~\ref{fig:delta_chi_onescale}, we compare the contribution to $\Delta\chi^2$ from different blocks of the covariance matrix (the full covariance, $BBBB$, $TBTB$ and $EBEB$ blocks) in each panel. In Fig.~\ref{fig:delta_chi_scales}, instead, we compare $\Delta\chi^2$ for different choices of $k_{\rm p}$ that can be probed by CMB observations, using the full covariance matrix.  
The white area in both figures represents the portion of parameter space excluded by current upper limits on the tensor-to-scalar ratio by \textit{Planck} and BICEP/Keck data at each scale \cite{Tristram:2020wbi, Tristram:2021tvh, BICEP:2021xfz, Campeti:2022vom, Hamann:2022apw}. Note that we assume $r_{*}=5r_{\rm vac}$ for all $k_{\rm p}$ choices, except for $k_{\rm p}=5\times10^{-3}\,\mathrm{Mpc}^{-1}$ for which we assume $r_{*}=r_{\rm vac}$, following Ref.~\cite{Ishiwata:2021yne} (section \ref{sec:theory}). In Figs.~\ref{fig:plot_raphael}, \ref{fig:plot_raphael_TB} and \ref{fig:plot_raphael_EB}, we also show the $BB$, $|TB|$ and $|EB|$ power spectra, respectively, used in Fig.~\ref{fig:delta_chi_scales}.

We find that the contribution to the total $\Delta\chi^2$ is entirely dominated by $BB$: {\sl LiteBIRD} will be able to  exclude with high significance the Starobinsky model if the tensor fluctuations arise from axion-SU(2) inflation and if the typical features of this latter model occur at scales observable by the CMB. Specifically, the significance is higher than $8\,\sigma$ for parameter close to the current upper limits on axion-SU(2) parameters, while $TB$ and $EB$ remain under $1\,\sigma$ significance in all cases considered, making it impossible for {\sl LiteBIRD} to detect parity-violation generated by the axion-SU(2) model. Furthermore, as noted in Ref.~\cite{Thorne:2017jft}, the $TB$ contribution dominates over the $EB$ one, making $\Delta\chi^2$ larger by a factor $\sim\mathcal{O}(10^2)$. 

Comparing the upper left and right panels of Fig.~\ref{fig:delta_chi_onescale}, we find that, for a given parameter choice, using only the $BBBB$ block results in higher significance compared to the full covariance matrix. This is because differences between the fiducial and axion-SU(2) model exist only in the $BB$ spectrum and marginally in $TB$ and $EB$. Therefore, using the full covariance just increases the number of degrees of freedom, diluting the information content.   
Although the inference of parameters by extracting a single block from the full covariance matrix is not strictly correct, it is still useful in this context to understand in an approximate fashion the contribution of different angular power spectra to the final {\sl LiteBIRD} sensitivity.

Our analysis can be improved. Since we are using the same sky mask, which was optimized for $B$ modes (see section \ref{sec:method}), for both temperature and polarization maps to compute the full covariance, the importance of the $TB$ spectrum in the analysis could further increase if we optimized masks to temperature field with a larger sky fraction.   

In summary, we conclude that {\sl LiteBIRD} will be able to  exclude with high significance tensor fluctuations produced by vacuum fluctuations (specifically, in the Starobinsky model)  if the GW background has been produced by the axion-SU(2) mechanism and the feature is sourced at the CMB scales. We also find that the $BB$ spectrum will dominate the discriminating power of {\sl LiteBIRD}, while $TB$ and $EB$ correlations would be of secondary importance. In the axion-SU(2) model, in fact, the production of detectable parity-violating correlations would imply an overproduction of $B$ modes, which would conflict with current observational limits on tensor modes.
Since the tensor power spectrum log-normal template in Eq.~\ref{eq:template} can include several different shapes, we can reframe our conclusions on parity-violating correlations $TB$ and $EB$ in a more general way. Based on our results in Fig.~\ref{fig:delta_chi_onescale}, we argue that \textit{LiteBIRD} will not be able to exclude the fiducial Starobinsky model using exclusively $TB$ and $EB$ correlations for all models producing fully or partially chiral GWs, as long as in the parameter space of interest: $(i)$ the tensor power spectrum shape is ``nested'' in the log-normal template;  $(ii)$ sourced scalar modes are negligible, or, in other words, the observational bounds on $TT$, $TE$ and $EE$ as well as on non-Gaussianities of the scalar perturbations, are satisfied; $(iii)$ the maximum allowed ratio between sourced and vacuum-produced tensor modes is $\lesssim O(10)$ at reionization bump scales and $\lesssim O(1)$ at recombination bump scales.  
We also note that higher-order correlations (such as bispectra, trispectra etc.) in the CMB are expected to be powerful probes to investigate parity-violation during inflation \cite{Gerbino:2016mqb}. A forecast using these statistics will be the subject of a future paper of the \textit{LiteBIRD} collaboration.

\section{Conclusions}\label{sec:conclusions}

Enhanced primordial gravitational waves from gauge fields during inflation represent a new paradigm of the primordial universe that has been extensively studied in the literature~\cite{Maleknejad:2012fw,Komatsu:2022nvu}. 
In this paper, we used realistic simulations to show that, thanks to the access to the reionization bump scales provided by a full-sky CMB mission, {\sl LiteBIRD} can provide significant help in our effort to distinguish between inflation models, more specifically in excluding the production of the stochastic GW background within the standard vacuum fluctuations paradigm in favor of production by matter sources and vice versa. We also presented expected constraints on the model parameters of an SU(2) model with a characteristic ``bump-like'' feature in the reionization bump. In this case, LiteBIRD will be able to obtain a two-sided 95\,\% C.L. confidence interval on the bump feature amplitude $r_{*}$.

The SU(2) model, in contrast to the standard inflationary scenario, also predicts parity-violating correlations in the CMB. We include the contribution of $TB$ and $EB$ in the full covariance matrix and assess the ability of {\sl LiteBIRD} to disentangle standard single-field slow-roll (specifically the Starobinsky model, a reference target for {\sl LiteBIRD}) from the axion-SU(2) model, finding that this experiment will be able to distinguish them with high significance for a wide range of model parameters. We also find that the discriminating power of {\sl LiteBIRD} will be determined mainly by $BB$, with $TB$ and $EB$ giving negligible contributions in almost all the allowed parameter space. Detecting the parity-violating signal in $TB$ and $EB$ from the axion-SU(2) model remains out of sight for {\sl LiteBIRD}.

We enforced bounds on sourced-to-vacuum tensor perturbations from the backreaction effect of spin-2 particle production on the background fields and from the non-Gaussian contribution to the scalar curvature perturbations produced at second order by sourced tensor modes, in order for the analytical template (Eq.~\ref{eq:template}) to remain valid. However, we stress that lattice simulations \cite{Caravano:2021bfn, Caravano:2022epk} can be used to relax these bounds and study the system also in the strong backreaction regime.

In this paper, we used the power spectrum as our main observable, taking advantage of the strong scale dependence of the tensor spectrum and the presence of parity-violating correlations predicted in axion-SU(2) models. However, additional information is available in the bispectrum and higher order correlation functions, potentially allowing to distinguish vacuum produced from sourced tensors when the power spectrum does not present significant features at CMB scales \cite{Namba:2015gja, Shiraishi:2016yun, Agrawal:2017awz,Agrawal:2018mrg}. We plan to explore this possibility in a dedicated future {\sl LiteBIRD} paper.

The axion-SU(2) model can also produce a signal directly detectable in the interferometer band, both in intensity and circular polarization of GWs \cite{Thorne:2017jft, Cai:2023ywp}. This opens up a new window in the GW spectrum to detect the parity-violating signature of this model. We also note that GW background anisotropies are a powerful probe of primordial parity-violation and non-Gaussianities in both PTA \cite{Kato:2015bye, Belgacem:2020nda, Tasinato:2023zcg, Fu:2023aab}
and interferometers \cite{Bartolo:2019oiq, Bartolo:2019yeu, LISACosmologyWorkingGroup:2022kbp}.

Finally, we mention the connection between this model and recent cosmic birefringence measurements. It has been shown in Ref.~\cite{Fujita:2022qlk} that the large $EB$ signal recently observed \cite{Minami:2020odp, Diego-Palazuelos:2022dsq, Eskilt:2022wav, Eskilt:2022cff} cannot be explained by a parity-violating GW background from the axion-SU(2) model, due to the simultaneous amplification of the $BB$ power spectrum that violates current upper bounds on the tensor-to-scalar ratio. If any signal is present in $EB$ from primordial parity-violation due to matter sources during inflation, it is expected to be subdominant with respect to the observed cosmic birefringence signal.

\acknowledgments

The authors thank José Luis Bernal, Tomohiro Fujita and Martin Reinecke for useful and stimulating discussion.
This work was supported in part by the Deutsche Forschungsgemeinschaft (DFG, German Research Foundation) under Germany's Excellence Strategy - EXC-2094 - 390783311.  This work has also received funding from the European Union's Horizon 2020 research and innovation programme under the Marie Sk\l odowska-Curie grant agreement No.~101007633.
This work is supported in Japan by ISAS/JAXA for Pre-Phase A2 studies, by the acceleration program of JAXA research and development directorate, by the World Premier International Research Center Initiative (WPI) of MEXT, by the JSPS Core-to-Core Program of A. Advanced Research Networks, and by JSPS KAKENHI Grant Numbers JP15H05891, JP17H01115, and JP17H01125.
The Canadian contribution is supported by the Canadian Space Agency.
The French \textit{LiteBIRD} phase A contribution is supported by the Centre National d’Etudes Spatiale (CNES), by the Centre National de la Recherche Scientifique (CNRS), and by the Commissariat à l’Energie Atomique (CEA).
The German participation in \textit{LiteBIRD} is supported in part by the Excellence Cluster ORIGINS, which is funded by the Deutsche Forschungsgemeinschaft (DFG, German Research Foundation) under Germany’s Excellence Strategy (Grant No. EXC-2094 - 390783311).
The Italian \textit{LiteBIRD} phase A contribution is supported by the Italian Space Agency (ASI Grants No. 2020-9-HH.0 and 2016-24-H.1-2018), the National Institute for Nuclear Physics (INFN) and the National Institute for Astrophysics (INAF).
Norwegian participation in \textit{LiteBIRD} is supported by the Research Council of Norway (Grant No. 263011) and has received funding from the European Research Council (ERC) under the Horizon 2020 Research and Innovation Programme (Grant agreement No. 772253 and 819478).
The Spanish \textit{LiteBIRD} phase A contribution is supported by the Spanish Agencia Estatal de Investigación (AEI), project refs. PID2019-110610RB-C21,  PID2020-120514GB-I00, ProID2020010108 and ICTP20210008.
Funds that support contributions from Sweden come from the Swedish National Space Agency (SNSA/Rymdstyrelsen) and the Swedish Research Council (Reg. no. 2019-03959).
The US contribution is supported by NASA grant no. 80NSSC18K0132.
The numerical analyses in this work have been supported by the Max Planck Computing and Data Facility (MPCDF) computer clusters \textit{Cobra}, \textit{Freya} and \textit{Raven}.

\bibliographystyle{JHEP}
\bibliography{biblio}
\end{document}